\documentclass[journal,10pt]{IEEEtran}
\usepackage{graphicx}
\usepackage{caption}
\usepackage{array}
\usepackage{amsfonts}
\usepackage{setspace}
\usepackage{amsmath}  
\usepackage{multirow}
\usepackage{mathtools}
\usepackage{epstopdf}
\usepackage{xcolor}
\usepackage[T1]{fontenc}
\usepackage{upquote}
\usepackage{amssymb}
 \usepackage{algorithm}
 \usepackage{algorithmic}

\usepackage[labelsep=period]{caption}
\newcounter{x}

\usepackage{cite}
\newtheorem{theorem}{Theorem}

\newtheorem{lemma}{Lemma}

\usepackage{algorithmic}
\usepackage{array}
\usepackage{fixltx2e}
\begin{document}

\title{Optimal Downlink Transmission for Cell-Free SWIPT Massive MIMO Systems with Active Eavesdropping}
%
\author{Mahmoud~Alageli,
        Aissa~Ikhlef,~\IEEEmembership{Senior Member,~IEEE}, Fahad Alsifiany, Mohammed A. M. Abdullah,~\IEEEmembership{Member,~IEEE}, Gaojie Chen,~\IEEEmembership{Senior Member,~IEEE}
        and~Jonathon~Chambers,~\IEEEmembership{Fellow,~IEEE}
%
\thanks{This work was supported by EPSRC grant number EP/R006377/1 (``M3NETs'').} 
\thanks{M. Alageli and F. Alsifiany are with the Intelligent Sensing and Communications Group, Newcastle University, NE1 7RU, UK. (e-mail: \{m.m.a.alageli1,~f.a.n.alsifiany2\}@ncl.ac.uk).}
\thanks{A. Ikhlef is with the Department of Engineering,
Durham University, Durham, DH1 3LE, UK. (e-mail:
aissa.ikhlef@durham.ac.uk).}
\thanks{M. A. M. Abdullah, G. Chen, and J. A. Chambers are with Department of Engineering, University of Leicester,
Leicester, U.K., LE1 7RH (email: \{m.abdullah,~gaojie.chen\}@leicester.ac.uk). Mohammed A. M. Abdullah is also a staff member with the Computer and Information Engineering Department, Ninevah University, Mosul, 41002, Iraq (email: mohammed.abdulmuttaleb@uoninevah@edu.iq). J. A. Chambers is also with the College of Automation, Harbin Engineering University, China.}

}
\maketitle
\begin{abstract} 
This paper considers secure simultaneous wireless information and power transfer (SWIPT) in cell-free massive multiple-input multiple-output (MIMO) systems. The system consists of a large number of randomly (Poisson-distributed) located access points (APs) serving multiple information users (IUs) and an information-untrusted dual-antenna active energy harvester (EH). The active EH uses one antenna to legitimately harvest energy and the other antenna to eavesdrop information. The APs are networked by a centralized infinite backhaul which allows the APs to synchronize and cooperate via a central processing unit (CPU). Closed-form expressions for the average harvested energy (AHE) and a tight lower bound on the ergodic secrecy rate (ESR) are derived. The obtained lower bound on the ESR takes into account the IUs' knowledge attained by downlink effective precoded-channel training. Since the transmit power constraint is per AP, the ESR is nonlinear in terms of the transmit power elements of the APs and that imposes new challenges in formulating a convex power control problem for the downlink transmission. To deal with these nonlinearities, a new method of balancing the transmit power among the APs via relaxed semidefinite programming (SDP) which is proved to be rank-one globally optimal is derived. A fair comparison between the proposed cell-free and the colocated massive MIMO systems shows that the cell-free MIMO outperforms the colocated MIMO over the interval in which the AHE constraint is low and vice versa. Also, the cell-free MIMO is found to be more immune to the increase in the active eavesdropping power than the colocated MIMO. 
\end{abstract}

\IEEEpeerreviewmaketitle
\section{Introduction} \label{Sec_Intro}
In contrast to multi-cell massive multiple-input multiple-output (MIMO) systems in which the users in each cell (of a confined area) are served by an array of colocated antennas, cell-free massive MIMO is an architecture in which the users over a large area are served by a large number of distributed antennas (access points (APs)) \cite{interdonato2018ubiquitous}. Given the provision of backhaul phase-coherent cooperation between the APs \cite{7827017,7468528,6963798}, the distributed deployment of the APs offers many advantages such as: eliminating the correlation between the transmitting antennas, the ability to overcome deep shadow fading, and more importantly, the large freedom in balancing the simultaneous transmissions of information, jamming and energy signals.   %

In massive MIMO systems, the asymptotic orthogonality between independent users' channels makes downlink transmission very robust against passive eavesdropping attacks \cite{7120012}. Therefore, the active eavesdropping attack in massive MIMO systems (which introduces correlation between the estimated channels of both the attacker and the attacked user) is relevant. Active information-eavesdropping relies on attacking the uplink channel estimation phase by sending an identical training sequence as the legitimate information user (IU), such that the estimated IU's channel is correlated with the channel of the attacking eavesdropper (EV). Therefore, the active EV benefits from the downlink transmission which is beamformed based on the estimated IU's channel \cite{7120012,7472059}.    

The broadcast nature of the wireless channel imposes challenges in securing wireless communication systems, particularly, in the presence of adversarial EVs \cite{mukherjee2014principles}. One example of such systems is simultaneous wireless information and power transfer (SWIPT) systems that comprise information-untrusted EHs. The secrecy issue in SWIPT massive MIMO systems, particularly under active attack, has previously lacked in-depth study in the literature. The main body of research concerning the secrecy problems in SWIPT systems has considered the colocated massive MIMO architecture \cite{zhang2016large,zhu2017wireless,wang2016secrecy,zhu2016wireless,8334259,alageli2018swipt}. The large dimensionality of transmit antennas in massive MIMO systems allows the use of random matrix theory to simplify the system design and performance analysis. Moreover, the asymptotic orthogonality between independent users' channels encourages the use of artificial noise (AN) jamming against any potential information eavesdropping.  In \cite{zhang2016large}, an asymptotic expression for the ergodic secrecy rate (ESR) of one IU and one passive information-untrusted energy harvester (EH) (both have multiple antennas) is derived in terms of the covariance matrix of the downlink signal vector. This asymptotic ESR is maximized by optimizing the covariance matrix subject to some average harvested energy (AHE) constraints. The AN jamming can be deployed in the downlink transmission phase to provide direct power transfer and to degrade the information signal quality at the EHs \cite{zhu2017wireless}. In \cite{8334259}, the use of AN is extended for both the downlink training and payload data transmission phases to further degrade the eavesdropping capabilities of the information EV. The authors in \cite{alageli2018swipt} considered joint enhancement of the secrecy and power transfer in the presence of an active dual-antenna information-untrusted EH. Asymptotic expressions for a lower bound on the ESR and the AHE are derived. Then, these results are used to optimize the power allocation for the downlink SWIPT transmission. Throughout the literature, much of the research regarding optimizing the performance of cell-free MIMO systems deals with the spectral efficiency \cite[and the references therein]{7827017}, the energy efficiency \cite{7900388,zhang2017spectral,8281464,8097026}, and the secrecy rate of wire-taped systems \cite{8360138}. 

This paper investigates the design and the performance evaluation of SWIPT in cell-free massive MIMO, particularly, the secrecy of the information transmission under an active attack from a dual-antenna information-untrusted EH. From the service provider (cooperative APs) point of view, the dual-antenna active EH's request for service equivalently appears as a separate legitimate EH using a training power $\phi P_{E}$ (where $0<\phi<1$ and $P_{E}$ is the total available training power) via the energy harvesting antenna, and illegitimate active EV attacking a certain IU with training power $(1-\phi) P_E$. However, the cooperative APs can rely on their large dimensionality to monitor the levels of training powers, therefore, they can blame the legitimate EH for the active attack. Upon the detection of the active attack, the cooperative APs have no option but to deal with this attack, and only two possible actions might be taken: 1) Dropping the IU under attack from service, i.e., stop sending information to the IU being attacked. With an exception for IUs receiving information with a high degree of importance, such an action seems impractical. Therefore, there is no secrecy design for the downlink transmission; 2) Dealing with the case by optimising the secrecy of the downlink transmission. Taking this action is useful and practical, particularly with the advantage of the large dimensionality of the APs. 

\textit{Contributions}: We are motivated by the lack of literature on the security of cell-free MIMO systems to provide a new globally optimal solution to the problem of joint power and data transfer in a cell-free massive MIMO system. The proposed system established by a large number of randomly (Poisson-distributed) located APs which cooperate via a central processing unit (CPU). The communication links between the APs and the IUs are vulnerable to be wire-tapped by an information-untrusted dual-antenna active EH. Since the transmit power constraint is per AP, the secrecy rate is nonlinear in terms of the transmit power elements of the APs and that imposes new challenges in formulating a convex power control problem for the downlink transmission. The main contributions of our work are: 1) To jointly improve the ESR and the AHE (of the legitimate EH), we propose optimized downlink transmissions of three different signals: information, AN and energy signals beamformed towards the IUs, legitimate and illegitimate antennas of the EH, respectively; 2) We derive closed-form expressions for the AHE and a tight lower bound on the ESR. The derived expressions are deterministic at the CPU and take into account the IUs' knowledge attained by downlink effective precoded-channel training; 3) Knowing that the ESR is nonlinear in terms of the transmit power elements of the APs, a new globally optimal iterative method for cooperatively balancing the transmit powers at the APs via relaxed semidefinite programming (SDP) is derived; 4) We provide a proof for the rank-one global optimality of our SDP solution (Theorem \ref{theorem_3}) and the convergence of our iterative SDP problem (Subsection \ref{converg}); 5) Finally, a fair performance comparison between the proposed cell-free and colocated massive MIMO systems is performed. The comparison shows informative results of the secrecy performance with respect to the active eavesdropping training power and the range of the AHE constraint values. 

\textit{Related Work:} To the best of the authors' knowledge, the secrecy performance in cell-free massive MIMO systems has only been studied in \cite{8360138} where the focus was on maximizing the secrecy rate of a given IU when being attacked by an active EV under constraints on the individual rates of all IUs. We can compare the work in this paper to the work in \cite{8360138} from two perspectives: 1) From system and signal design perspectives, our work considers the worst-case SWIPT problem by optimizing three different downlink signals: information, AN and energy signals beamformed towards the IUs, legitimate and illegitimate antennas of the dual-antenna EH, respectively; while work in \cite{8360138} considers the secrecy problem of a certain IU by optimizing the downlink information signals (no jamming or power transfer are considered); 2) From a problem-solving perspective, the employed lower bound on the secrecy rate in \cite{8360138} imposes constraints on the domain of the linear programming (LP) optimization variables (the allocated power of the downlink information vectors) \cite[(23)]{8360138}, i.e., the values of allocated power vectors are feasible on a sub-region of $\mathcal{R}_{+}^{N}$, $N$ is the total number of APs. Since the update in the proposed iterative algorithm does not include the power vector of the considered IU, the obtained solution is locally optimal, or at least, the globally optimal solution is not guaranteed. In contrast, in our work, both the objective function and constraints of the SDP formulation are differentiable and there are no constraints on the domain of the optimization variables which implies the satisfaction of Slater's condition. Therefore, by proving the optimal rank requirements (please see Theorem \ref{theorem_3} and its proof) and the convergence of the employed iterative problem (please see Subsection \ref{converg}), we claim the global optimality of our solution. In our early work in \cite{alageli2018swipt}, an active dual-antenna information-untrusted EH (equivalent to the proposed EH in this paper) has been considered for a colocated SWIPT massive MIMO system. However, considering such a secrecy problem for cell-free massive MIMO will result in a non-linear objective function in terms of the allocated power elements at the APs. Inevitably, this problem can not be solved by the LP method used for a colocated massive MIMO in \cite{alageli2018swipt}, and this leads to a completely different SDP optimization challenge.  

\textit{Notation:} For referencing convenience, the notations used in this paper are listed in Table I at the top of the next page. 
 
 \begin{center}
 \begin{table}[h]\label{tab:table1}
\centering\footnotesize
\renewcommand{\arraystretch}{1.3}
\captionsetup{justification=centering, labelsep=newline}
\caption {\scshape List of Notations}
    \begin{tabular}{ |m{1.2cm}|m{6.6cm}| }
    \hline
    Notation & Description   \\ \hline
    $\boldsymbol{a}$, $\boldsymbol{A}$& vectors and matrices are denoted by boldface lowercase and boldface uppercase letters, respectively\\ \hline
    $\boldsymbol{I}_{N}$ & denotes the $N\times N$ identity matrix  \\ \hline
    $\text{diag}(\boldsymbol{s})$ &a matrix whose diagonal entries are the entries of vector $\boldsymbol{s}$ and zeros elsewhere \\ \hline
    $\text{diag}(\boldsymbol{S})$ & a column vector whose entries are the diagonal entries of matrix $\boldsymbol{S}$\\ \hline
    $\boldsymbol{S}\succeq 0$& indicates that $\boldsymbol{S}$ is a positive semidefinite matrix\\ \hline 
    $(\cdot)^T~\text{and}$ $(\cdot)^H$  & the transpose and the conjugate transpose, respectively \\ \hline
    $\text{tr}(\cdot)~\text{and}$ $\text{log}_{2}(\cdot)$ & the trace of a matrix and logarithm to base 2, respectively \\ \hline
    $|\cdot|~\text{and}$ $\|\cdot\|$ & the absolute value of scalars and the Euclidean norm, respectively \\ \hline 
   $\mathcal{R},~\mathcal{R}_{+}^{n}$, $\mathcal{S}_{+}^{ n}$ and $\mathcal{C}^{m\times n}$& sets of real numbers, nonnegative real numbers, symmetric positive semidefinite $n\times n$ real matrices and complex $m\times n$ matrices, respectively \\ \hline
   $\mathcal{CN}(\boldsymbol{0},\boldsymbol{\Sigma})$& circularly symmetric complex Gaussian distribution of  a random vector with zero mean and covariance matrix $\boldsymbol{\Sigma}$\\ \hline
   $\text{cov}(x,y)$ $\text{and}~\text{var}(x)$ & the covariance between the random variables (RVs) $x$ and $y$, and the variance of $x$, respectively\\ \hline
   $\{\boldsymbol{a}_{n}\}~\text{and}$ $\{a_{m,n}\}_{m}$& a set of all vectors indexed by $n$ and a set of all scalars indexed by $m$, respectively \\\hline
   $[\boldsymbol{a}]_{n}~\text{and}$ $[\boldsymbol{A}]_{n,m}$ & the $n$th entry of $\boldsymbol{a}$ and the $(n,m)$th entry of $\boldsymbol{A}$, respectively\\ \hline
   $\boldsymbol{B}=\text{null}\left(\boldsymbol{A}\right)$ & means $\boldsymbol{A}\boldsymbol{B}=\boldsymbol{0}$ and $\boldsymbol{B}\boldsymbol{B}^{H}=\boldsymbol{I}$\\ \hline
   $[x]^{+}$ & is equivalent to $\max\;(x,0)$ \\ \hline
    \end{tabular}
\end{table}
\end{center}

\section{System Model}\label{Sec_SysMod}  
\begin{figure}[t]
\begin{center}
\captionsetup{justification=centering}
\includegraphics[width=.4\textwidth,trim = 0cm .0cm 0cm .0cm, clip]{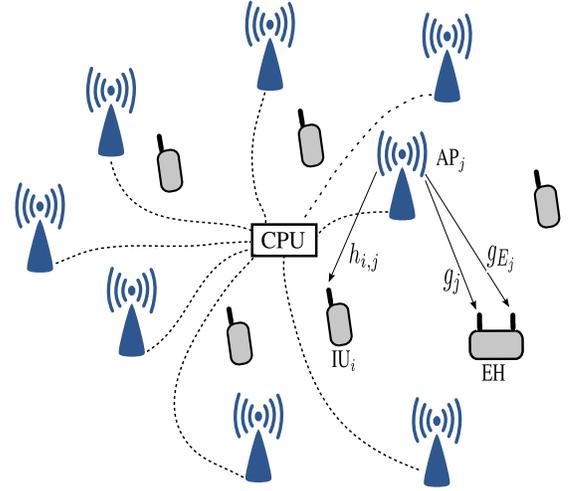}
\caption{An illustration of the proposed SWIPT cell-free massive MIMO system, only a small number of APs is illustrated for clarity.}
\label{fig00}
\end{center}
\end{figure}

As illustrated in Fig. \ref{fig00}, we consider the downlink of a cell-free massive MIMO system consisting of a large number of APs which are randomly located on a two dimensional Euclidean area $\text{A}_a$ based on an homogeneous Poisson point process (PPP) $\Phi_a$ with an intensity $\lambda_a$; $M$ single antenna IUs interested in information decoding, $\{\text{IU}_{i}\}$, $i=1,2,..., M$; and an active information-untrusted EH, equipped with two antennas, where one antenna is used to legitimately harvest energy, while the other antenna is used to illegitimately and actively eavesdrop and decode an information signal intended for a certain IU, $\text{IU}_{k}$, $k\in \{1,2,..., M\}$. Unless otherwise stated, the IUs and the EH are randomly located on a two dimensional Euclidean area $\text{A}_{u} < \text{A}_{a}$\footnote{Since each user (IU or EH) is dominantly served by a subset of the APs. Therefore, the assumption $\text{A}_{u} < \text{A}_{a}$ introduces an overlap between the dominant AP groups serving different users. From the secure SWIPT design point of view, this case is more severe than the case when the users are widely apart, i.e., $\text{A}_{u} = \text{A}_{a}$.}. The origins of both $\text{A}_{u}$ and $\text{A}_{a}$ coincide. The APs are networked by a centralized infinite backhaul which allows them to synchronize and cooperate via a CPU. 

Let $\{\text{AP}_{1},\;\dots,\text{AP}_{N}\}$ be the set of the adopted realization of APs. ${\fontsize{9.29}{11}\boldsymbol{h}_{i}=[h_{i,1},\;\dots,h_{i,N}]^{T}=\boldsymbol{\Gamma}_{i}^{\frac{1}{2}}\boldsymbol{\bar{h}}_{i}}$ denotes the uplink channel vector between $\text{IU}_{i}$ and the set of APs, where ${\fontsize{9.29}{11}\boldsymbol{\bar{h}}_{i}\sim\mathcal{CN}(\boldsymbol{0},\boldsymbol{I}_{N})}$ is the small-scale fading vector and ${\fontsize{9.29}{11}\boldsymbol{\Gamma}_{i}=\text{diag}([\gamma_{i,1},\;\dots,\gamma_{i,N}])}$, $\gamma_{i,j}$ is the large-scale fading coefficient of the channel between $\text{IU}_{i}$ and $\text{AP}_{j}$. ${\fontsize{9.29}{11}\boldsymbol{g}=[g_{1},\;\dots,g_{N}]^{T}=\boldsymbol{\Gamma}^{\frac{1}{2}}\boldsymbol{\bar{g}}}$ and ${\fontsize{9.29}{11}\boldsymbol{g}_{E}=[g_{E_1},\;\dots,g_{E_N}]^{T}=\boldsymbol{\Gamma}^{\frac{1}{2}}\boldsymbol{\bar{g}}_{E}}$ denote the uplink channel vectors between the legitimate and the illegitimate (eavesdropping) antennas of the EH and the set of APs, respectively, where ${\fontsize{9.29}{11}\boldsymbol{\bar{g}}}=[\bar{g}_{1},\;\dots,\bar{g}_{N}]^{T},~{\fontsize{9.29}{11}\boldsymbol{\bar{g}}_{E}}=[\bar{g}_{E_1},\;\dots,\bar{g}_{E_N}]^{T}\sim\mathcal{CN}(\boldsymbol{0},\boldsymbol{I}_{N})$ are independent, uncorrelated small-scale fading vectors. ${\fontsize{9.29}{11}\boldsymbol{\Gamma}=\text{diag}([\gamma_{1},\;\dots,\gamma_{N}])}$ where $\gamma_{j}$ is the large-scale fading coefficient of the channel between the EH and $\text{AP}_{j}$. The large-scale fading coefficients $\{\gamma_{i,j},~\gamma_{j}\}$ change very slowly compared to the small-scale fading coefficients, therefore, we assume that $\{\gamma_{i,j},~\gamma_{j}\}$ are perfectly known at the APs \cite{6746659}.  
\subsection{Uplink Channel Estimation}\label{UP_Training}   
The user small-fading channels manifest block fading, i.e., they remain constant over one time block, but change independently from one block to another. Each time block is divided into three time slots of lengths: $\tau$ transmission samples for uplink training, $\tau_d$ transmission samples for downlink training and $\tau_s$ samples for downlink data transmission. Without loss of generality, we assume a unit time slot for the downlink data transmission $\tau_{s}T_{s}= 1s$, where $T_s$ is the duration of the transmitted data symbol \cite{zhang2016large,6728676}. During the uplink training phase, a training sequence is sent from each IU with an average power $P_I$. Pessimistically, we assume that the EH has the potential to acquire the training sequence of a certain IU (made possible by overhearing the leaking electromagnetic signalling between the APs and the IUs \cite{young2016information}). Therefore, the EH sends a copy of the training sequence of the attacked IU, $\text{IU}_{k},~k\in\{1,2,~\dots,M\}$, via its eavesdropping antenna using part of its total average power $\phi P_{E}$, $0<\phi<1$, such that the cooperative APs estimate the uplink composite channel coefficients of both $\text{IU}_{k}$ and the eavesdropping antenna of the EH. Consequently, the estimated channel of $\text{IU}_{k}$ will be corrupted and correlated with the illegitimate channel of the EH \cite{7120012,zhou2012pilot}. The remaining training power $(1-\phi) P_{E}$ is used for transmitting the legitimate uplink training sequence via the energy harvesting antenna. The uplink training sequences of the IUs and legitimate EH are assumed to be orthogonal. The signal at the APs received across $\tau$ training transmissions is
\begin{equation}\begin{aligned}\label{eeq1} 
 &\boldsymbol{Y} =\\
 &\sum_{i=1}^{M}\sqrt{P_{I}}\;\boldsymbol{h}_{i}\;\boldsymbol{\psi}_{i}^{T}+\sqrt{\phi P_{E}}\boldsymbol{g}_{E}\boldsymbol{\psi}_{k}^{T}+\sqrt{(1-\phi)P_{E}}\boldsymbol{g}\boldsymbol{\psi}_{E}^{T}+\boldsymbol{N},
\end{aligned}\end{equation} 
 
  \noindent
where $\boldsymbol{N} \in \mathcal{C}^{N \times \tau}$ is the additive noise matrix with entries following the distribution $\mathcal{CN}(0,\sigma_{n}^{2})$. $k$ is the index of the attacked IU, $\text{IU}_{k}$. $\boldsymbol{\psi}_{i},~\boldsymbol{\psi}_{k},~\boldsymbol{\psi}_{E}\in \mathcal{C}^{\tau\times 1}$ are the uplink training sequences of $\text{IU}_{i}$, the IU under attack, $\text{IU}_{k}$, and the legitimate antenna of the EH, respectively. $\boldsymbol{\psi}_{i}^{H}\boldsymbol{\psi}_{j\neq i},\;\boldsymbol{\psi}_{i}^{H}\boldsymbol{\psi}_{E}=0$; and $\boldsymbol{\psi}_{i}^{H}\boldsymbol{\psi}_{i},\;\boldsymbol{\psi}_{E}^{H}\boldsymbol{\psi}_{E}=\tau $. We assume centralized channel estimation via the CPU. Given that $\text{IU}_{k}$ is the attacked IU, the minimum mean square error (MMSE) estimate of $\boldsymbol{h}_{i}$, $\boldsymbol{\hat{h}}_{i}=[\hat{h}_{i,1},\;\dots,\hat{h}_{i,N}]^{T}$, and of $\boldsymbol{g}$, $\boldsymbol{\hat{g}}=[\hat{g}_{1},\;\dots,\hat{g}_{N}]^{T}$, are given as 
\begin{subequations}\label{eeq2}
\begin{flalign}
&\boldsymbol{\hat{h}}_{i}=\boldsymbol{C}_{i}\boldsymbol{y}_{i},~\boldsymbol{C}_{i}=\sqrt{P_I}\boldsymbol{\Gamma}_{i}\left(\tau P_{I}\boldsymbol{\Gamma}_{i}+\delta_{ik}\;\tau\phi P_{E}\boldsymbol{\Gamma}+\sigma_{n}^{2}\boldsymbol{I}_{N}\right)^{-1},\label{eeq2a}\\
& \boldsymbol{y}_{i}=\boldsymbol{Y}\boldsymbol{\psi}_{i}^{*}=\tau \sqrt{P_{I}}\boldsymbol{h}_{i}+ \delta_{ik}\;\tau \sqrt{\phi P_{E}}\;\boldsymbol{g}_{E} + \boldsymbol{N}\boldsymbol{\psi}_{i}^{*},\label{eeq2c}\\
&\boldsymbol{\hat{g}}=\boldsymbol{C}\boldsymbol{y},~~\boldsymbol{C}=\sqrt{(1-\phi) P_{E}}\boldsymbol{\Gamma}~\left(\tau (1-\phi) P_{E}\boldsymbol{\Gamma}+\sigma_{n}^{2}\boldsymbol{I}_{N}\right)^{-1},\label{eeq2b}\\
& \boldsymbol{y}=\boldsymbol{Y}\boldsymbol{\psi}_{E}^{*}=\tau \sqrt{(1-\phi) P_{E}}\;\boldsymbol{g} + \boldsymbol{N}\boldsymbol{\psi}_{E}^{*},\label{eeq2d}
\end{flalign}\end{subequations}
\noindent
where $\delta_{ik}=1$ if $i=k$ (i.e., $\text{IU}_{i}$ is the attacked IU) and $\delta_{ik}=0$ if $i\neq k$. The covariance matrices $\mathbb{E}[\boldsymbol{\hat{h}}_{i}\boldsymbol{\hat{h}}_{i}^{H}]$ and $\mathbb{E}[\boldsymbol{\hat{g}}\boldsymbol{\hat{g}}^{H}]$ are equal to $\boldsymbol{R}_{i}=\tau \sqrt{P_I}\boldsymbol{\Gamma}_{i}\boldsymbol{C}_{i}$ and $\boldsymbol{R}=\tau \sqrt{(1-\phi) P_{E}}\boldsymbol{\Gamma}\boldsymbol{C}$, respectively. To emphasize whether $\text{IU}_{i}$ is being attacked or not, we use $\boldsymbol{R}_{i}$ to describe the covariance matrix of $\text{IU}_{i}$ if not being attacked and $\boldsymbol{\bar{R}}_{i}$ to describe the covariance matrix of $\text{IU}_{i}$ if being attacked. Both $\boldsymbol{R}_{i}$ and $\boldsymbol{\bar{R}}_{i}$ are calculated by the same aforementioned formula, but with $k\neq i$ for $\boldsymbol{R}_{i}$ and with $k=i$ for $\boldsymbol{\bar{R}}_{i}$. The results in \eqref{eeq2a} and \eqref{eeq2b} follow from standard channel estimation theory \cite{jagannatham2016noc,kay1993fundamentals}. Active eavesdropping attack detection and the identification of the attacked IU, $\text{IU}_{k}$, are possible and have been studied in \cite{tugnait2018pilot,tugnait2017detection,zhang2018detection}. Alternatively, the cooperative APs can exploit their large dimensionality to detect the active eavesdropping attack by monitoring the values of training powers which have been proven to be accurate as $N\to\infty$. The CPU can calculate the eavesdropping (illegitimate) and the legitimate training powers of the EH, $\phi P_{E}$ and $(1-\phi)P_{E}$, respectively, by using the following lemma\footnote{Since the cooperative APs are able to monitor the
changes in the training powers of the IUs and the EH using Lemma 1, we assume that the cooperative APs blame the information-untrusted EH for the active eavesdropping attack.} 

\begin{lemma}\label{lemma1c}
For a large density of APs as $\lambda_{a}\to\infty$, which leads to a large number of APs as $N\to\infty$, any illegitimate active training power can be identified and calculated as 

{\fontsize{9.55}{11}\begin{equation}\label{eeq3}
\frac{\boldsymbol{y}_{i}^{H}\boldsymbol{y}_{i}-\tau^{2}P_{I}\text{tr}\left(\boldsymbol{\Gamma}_{i}\right)-N\tau\sigma_{n}^{2}}{\tau^{2}\text{tr}\left(\boldsymbol{\Gamma}\right)}\stackrel{N\to\infty}{\to}\delta_{ik}~\phi P_{E} , 
\end{equation}}

\noindent
where $\text{IU}_{i}$ is under attack if $\delta_{ik}=1$, i.e., $k=i$, and $\text{IU}_{i}$ is not being attacked if $\delta_{ik}=0$, i.e., $k\neq i$. All the scalars, vector and matrices in the left-hand side of \eqref{eeq3} are deterministic at the CPU.  
\end{lemma}

\begin{IEEEproof}
See Appendix \ref{Appen_A-1}. 
\end{IEEEproof}

\subsection{Downlink Transmission}\label{Training}   
The APs cooperate via the CPU to control the power allocation of the downlink data, AN, and energy signal transmissions. From the service provider (cooperative APs) point of view, the EH's request for service equivalently appears to the cooperative APs as a separate legitimate EH which uses a training power $\phi P_E$ and illegitimate active eavesdropper attacking a certain IU, $\text{IU}_{k}$, with a training power $(1-\phi) P_E$. However, the CPU relies on the large dimensionality of the APs to monitor the levels of training powers, and based on Lemma \ref{lemma1c}, it can blame the legitimate EH for the active attack. Upon the detection of the active attack, the CPU has no option but to deal with this attack, and only two possible actions might be taken:
\begin{itemize}
\item Dropping the IU under attack from service, i.e., stop sending information to the IU being attacked. With an exception for IUs receiving information with a high degree of importance, such an action seems impractical. Therefore, there is no secrecy design for the downlink transmission.

\item Dealing with the case by optimizing the secrecy of downlink transmission (by employing controlled transmissions of information, jamming and energy signals). Taking this action is useful and practical, particularly with the advantage of the large number of randomly located APs. Compared to the case of collocated APs (conventional MIMO), the average path-loss from an AP to the active EH and the attacked IU varies from one AP to another. This property of randomly distributed APs would increase the efficiency of power control in tackling the active eavesdropping.       
\end{itemize}

Given that the $\text{IU}_{k}$ is the attacked IU, the APs employ the matched filter (MF) precoder to transmit the downlink signal vector
\begin{equation}\label{eeq3new}
\boldsymbol{x}_{k}=\sum_{i=1}^{M}\boldsymbol{w}_{i}q_{i}+\boldsymbol{\bar{w}}_{k}z+\boldsymbol{w}, 
\end{equation}
 
 \noindent
 where the $j$th entry of $\boldsymbol{x}_{k}$, $[\boldsymbol{x}_{k}]_{j}$, is the signal transmitted by $\text{AP}_{j}$, $\boldsymbol{w}_{i}q_{i}$ is the information signal vector directed towards $\text{IU}_{i}$, $\boldsymbol{\bar{w}}_{k}z$ is the AN signal vector directed towards the eavesdropping antenna of the EH, and $\boldsymbol{w}$ is the energy signal vector directed towards the legitimate antenna of the EH. $\{q_{i}\}$ and $z$ are the information signal symbols intended for $\{\text{IU}_{i}\}$ and the AN symbol, respectively, and they are mutually independent and follow the distribution $\mathcal{CN}(0,1)$. The MF beamforming vectors in \eqref{eeq3new} are defined as\footnote{Please note, due to active attack, the $\boldsymbol{\hat{h}}_{k}$ used to design $\boldsymbol{\bar{w}}_{k}$ in \eqref{eeq4b} is the estimate of the composite channel of both $\boldsymbol{h}_{k}$ and $\boldsymbol{g}_{E}$. By optimizing the per AP AN power factors $\{\bar{p}_{1},\;\dots,\;\bar{p}_{N}\}$, the AN power can be maximized at the EH and minimized at the $\text{IU}_k$.} 
\begin{subequations}\begin{flalign}
 &\boldsymbol{w}_{i}=\text{diag}\left(\boldsymbol{p}_{i}\right)\boldsymbol{\hat{h}}_{i}^{*},~\boldsymbol{p}_{i}=\left[\sqrt{p_{i,1}},\;\dots,\;\sqrt{p_{i,N}}\right]^{T},\label{eeq4a}\\
 &\boldsymbol{\bar{w}}_{k}=\text{diag}\left(\boldsymbol{\bar{p}}\right)\boldsymbol{\hat{h}}_{k}^{*},~\boldsymbol{\bar{p}}=\left[\sqrt{\bar{p}_{1}},\;\dots,\;\sqrt{\bar{p}_{N}}\right]^{T},\label{eeq4b}\\
 & \boldsymbol{w}=\text{diag}\left(\boldsymbol{p}\right)\boldsymbol{\hat{g}}^{*},~\boldsymbol{p}=\left[\sqrt{p_{1}},\;\dots,\;\sqrt{p_{N}}\right]^{T}.\label{eeq4c}
\end{flalign}\end{subequations} 

  \noindent 
For example, $|\left[\boldsymbol{w}_{i}\right]_{j}|^{2}=p_{i,j}|\hat{h}_{i,j}|^{2}$, $|\left[\boldsymbol{\bar{w}}_{k}\right]_{j}|^{2}=\bar{p}_{j}|\hat{h}_{k,j}|^{2}$ and $|\left[\boldsymbol{w}\right]_{j}|^{2}=p_{j}|\hat{g}_{j}|^{2}$ are the allocated powers at $\text{AP}_{j}$ for $\text{IU}_{i}$'s data, AN and energy signals, respectively. Power allocation is controlled via the factors $\{p_{i,j}\}$, $\{\bar{p}_{j}\}$ and $\{p_{j}\}$. Referring to \eqref{eeq2a} and \eqref{eeq4b}, it can be noticed that the received AN signal power at the eavesdropping antenna of the EH, $|\boldsymbol{g}_{E}^{T}\boldsymbol{\bar{w}}_{k}|^{2}$, is directly proportional to the eavesdropping training power, $\phi P_{E}$, i.e., the larger the eavesdropping training power, the larger the jamming received power by the EH. Therefore, although the AN is aligned to the $\text{IU}_{k}$'s estimated channel coefficients, the cooperative APs can improve the information secrecy by exploit the nature of the cell-free system -- in which $\text{IU}_{k}$ and the EH experience different path-losses to a single AP -- by optimizing the per AP per user power control.  
  
Given that $\text{IU}_{k}$ is the attacked IU. The received signals at $\text{IU}_{i}$, $y_{k,i}$; the legitimate antenna of the EH, $y_{k}$; and at the eavesdropping antenna of the EH, $y_{E_k}$, are          
%
%
\begin{subequations}\begin{flalign}
 &y_{k,i}=\boldsymbol{h}_{i}^{T}\boldsymbol{x}_{k}+n_{i},\label{eeq5a}\\
 &y_{k}=\boldsymbol{g}^{T}\boldsymbol{x}_{k}+\hat{n},\label{eeq5b}\\
 &y_{E_k}=\boldsymbol{g}_{E}^{T}\boldsymbol{x}_{k}+\bar{n}.\label{eeq5c}
\end{flalign}\end{subequations} 
\noindent
where $n_{i}$, $\hat{n}$ and $\bar{n}$ are zero mean $\sigma_{n}^{2}$ variance complex Gaussian noises at $\text{IU}_{i}$, the legitimate and eavesdropping antennas of the EH, respectively.
%
\subsection{Downlink Effective Precoded-Channel Estimation}\label{Dow_Training}   

With a large number of APs, the channel estimation at all IUs requires training sequences of a length $\geq N$ which is practically infeasible. Alternatively, we propose the estimation of the effective precoded-channels, $\{a_{i,i}=\boldsymbol{h}_{i}^{T}\boldsymbol{w}_{i}\}$ at the IUs\footnote{The EH has the potential to estimate the precoded channel for the attacked IU, $b_{k}=\boldsymbol{g}_{E}^{T}\boldsymbol{w}_{k}$, however, as will be seen in Subsection \ref{UPEH_rate}, the worst case in which the EH can perfectly estimate $b_{k}$ is assumed.}. The downlink estimation of the effective precoded-channels at the IUs requires $M$ orthogonal training sequences that can be of a finite length, $\geq M$. Therefore, such a downlink estimation is practically possible. Notice that $\text{IU}_{i}$ needs to estimate its effective precoded-channel $a_{i,i}$ which includes the values of power control factors $\{p_{i,j}\}$, $\{\bar{p}_{j}\}$ and $\{p_{j}\}$. Therefore the values of $\{p_{i,j}\}$, $\{\bar{p}_{j}\}$ and $\{p_{j}\}$ to be used for downlink data transmission are employed for downlink training. The cooperative APs transmit the downlink training signal matrix $\boldsymbol{X}_{d}=\sum_{i=1}^{M}\boldsymbol{w}_{i}~\boldsymbol{\psi}_{d_i}^{T}$, where $\{\boldsymbol{\psi}_{d_i}\}$ -- $\boldsymbol{\psi}_{d_i}^{H}\boldsymbol{\psi}_{d_i}=\tau_d$ and $\boldsymbol{\psi}_{d_i}^{H}\boldsymbol{\psi}_{d_{j\neq i}}=0$ -- are the downlink training sequences of the IUs\footnote{ The same training sequences could be used in the uplink and downlink.}. The received training signal vector at $\text{IU}_{i}$, $\boldsymbol{y}_{I_i}\in\mathcal{C}^{1\times \tau_d}$ is  
\begin{equation}
\boldsymbol{y}_{I_i}=\boldsymbol{h}_{i}^{T}\boldsymbol{X}_{d}+\boldsymbol{n}_{i}=\sum_{j=1}^{M}a_{i,j}\boldsymbol{\psi}_{d_j}^{T}+\boldsymbol{n}_{i},\label{eeq6a}\\
\end{equation}
   
\noindent
where $a_{i,j}=\boldsymbol{h}_{i}^{T}\boldsymbol{w}_{j}$ and $\boldsymbol{n}_{i}\sim \mathcal{CN}(0,\sigma_{n}^{2}\boldsymbol{I}_{\tau_d})$ is the noise vector at $\text{IU}_{i}$. First, let us examine the MMSE estimate of $a_{i,i}$ at $\text{IU}_{i}$ which can be calculated as \cite{jagannatham2016noc,kay1993fundamentals} 
\begin{equation}\begin{aligned}\label{eeq7}
\frac{\boldsymbol{p}_{i}^{T}\boldsymbol{\Gamma}_{i}\boldsymbol{R}_{i}\boldsymbol{p}_{i}}{\boldsymbol{p}_{i}^{T}\boldsymbol{\Gamma}_{i}\boldsymbol{R}_{i}\boldsymbol{p}_{i}+\tau_{d}\sigma_{n}^{2}}~y_{I_i},
\end{aligned}\end{equation}

\noindent
where $y_{I_i}=\boldsymbol{y}_{I_i}\boldsymbol{\psi}_{d_i}^{*}=\tau_d a_{i,i}+\boldsymbol{n}_{i}\boldsymbol{\psi}_{d_i}^{*}$. However, since the allocated power control factors in $\boldsymbol{p}_{i}$ are not available at $\text{IU}_{i}$, the calculation of \eqref{eeq7} is not possible, and instead, we assume that $\text{IU}_{i}$ performs a simple least square error (LSE) estimate of $a_{i,i}$, $\hat{a}_{i,i}$ which is given as
\begin{equation}\begin{aligned}\label{eeq8}
\hat{a}_{i,i}=\frac{y_{I_i}}{\tau_d}=a_{i,i}+\tilde{a}_{i,i},
\end{aligned}\end{equation} 

\noindent
where $\tilde{a}_{i,i}=\frac{\boldsymbol{n}_{i}\boldsymbol{\psi}_{d_i}^{*}}{\tau_d}$ is the estimation error which is statistically independent from the effective precoded channel $a_{i,i}$.

\section{Secrecy Analysis}\label{Sec_Analysis} 
\subsection{Lower Bound on the IU Rate}\label{LBIU_rate} 
The received signal at $\text{IU}_{i}$ given in \eqref{eeq5a} can be recast as follows
\begin{equation}\begin{aligned}\label{eeq10}
y_{k,i}&=a_{i,i} q_{i}+Z_{k,i}\\
&=\mathbb{E}\left[a_{i,i}|\hat{a}_{i,i}\right]q_{i}+\left(a_{i,i}-\mathbb{E}\left[a_{i,i}|\hat{a}_{i,i}\right]\right)q_{i}+Z_{k,i},
\end{aligned}\end{equation}

\noindent
where
\begin{equation}\begin{aligned}\label{eeq11}
Z_{k,i}=\sum_{j\neq i}a_{i,j}q_{j}+\boldsymbol{h}_{i}^{T}\left(\boldsymbol{\bar{w}}_{k}z+\boldsymbol{w}\right)+n_{i}.
\end{aligned}\end{equation}

\noindent
$\mathbb{E}\left[a_{i,i}|\hat{a}_{i,i}\right]q_{i}$ is the desired information signal received through a deterministic precoded channel $\mathbb{E}\left[a_{i,i}|\hat{a}_{i,i}\right]$, while $\left(a_{i,i}-\mathbb{E}\left[a_{i,i}|\hat{a}_{i,i}\right]\right)q_{i}$ is the desired information signal received through a non-deterministic precoded channel $a_{i,i}-\mathbb{E}\left[a_{i,i}|\hat{a}_{i,i}\right]$.  $\mathbb{E}\left[a_{i,i}|\hat{a}_{i,i}\right]q_{i}$ and $\left(a_{i,i}-\mathbb{E}\left[a_{i,i}|\hat{a}_{i,i}\right]\right)q_{i}$ are statistically dependent. $Z_{k,i}$ is the equivalent noise\footnote{$Z_{k,i}$ is considered as an equivalent noise since $Z_{k,i}$ and $\mathbb{E}\left[a_{i,i}|\hat{a}_{i,i}\right]q_{i}$ are independent and that follows since $\{q_{j}\}$, $z$, $n_{i}$ and $\boldsymbol{w}$ are statistically independent.} which accounts for inter user interference, energy signal interference and the thermal noise. Referring to \eqref{eeq8}, we can see that $a_{i,i}$ is explicitly decoupled and therefore $a_{i,i}$ and $\tilde{a}_{i,i}$ are uncorrelated and statistically independent. Since $\hat{a}_{i,i}$ is deterministic at $\text{IU}_{i}$, then
\begin{subequations}\begin{flalign}
 &\mathbb{E}\left[a_{i,i}|\hat{a}_{i,i}\right]\nonumber\\
 &=\mathbb{E}\left[\hat{a}_{i,i}|\hat{a}_{i,i}\right]+\mathbb{E}\left[\tilde{a}_{i,i}|\hat{a}_{i,i}\right]=\hat{a}_{i,i}+\mathbb{E}\left[\tilde{a}_{i,i}\right]=\hat{a}_{i,i},\label{eeqjk12}\\
&a_{i,i}-\mathbb{E}\left[a_{i,i}|\hat{a}_{i,i}\right]=\tilde{a}_{i,i}, 
 \end{flalign}\end{subequations}
 
 \noindent
 where $\mathbb{E}\left[\hat{a}_{i,i}|\hat{a}_{i,i}\right]=\hat{a}_{i,i}$ follows as an expectation over a deterministic value; $\mathbb{E}\left[\tilde{a}_{i,i}|\hat{a}_{i,i}\right]=\mathbb{E}\left[\tilde{a}_{i,i}\right]$ follows from the statistical independence between $\tilde{a}_{i,i}$ and $\hat{a}_{i,i}$; and $\mathbb{E}\left[\tilde{a}_{i,i}\right]=0$ follows since $\mathbb{E}\left[\boldsymbol{n}_{i}\boldsymbol{\psi}_{d_i}^{*}\right]=0$. Using the results in \cite[Theorem 1]{jose2011pilot} and in \cite[(22)]{interdonato2016much}, the downlink information rate at the attacked user $\text{IU}_{k}$, $R_{k}$ (given in \eqref{eeq13}) is achievable and forms a lower bound on the ergodic information rate 
 
\begin{equation}\begin{aligned}\label{eeq13}
R_{k}=\mathbb{E}\left\{\text{log}_{2}\left(1+\text{SINR}_{k}\right)\right\},
\end{aligned}\end{equation}
where
\begin{equation}\begin{aligned}\label{eeq14}
\text{SINR}_{k}&=\frac{\left|\hat{a}_{k,k}\right|^{2}}{\mathbb{E}\left[\left|a_{k,k}-\mathbb{E}\left[a_{k,k}|\hat{a}_{k,k}\right]\right|^{2}\right]+\mathbb{E}\left[\left|Z_{k,k}\right|^{2}\right]}\\
&=\frac{\left|\hat{a}_{k,k}\right|^{2}}{\text{var}\left(\tilde{a}_{k,k}\right)+\text{var}\left(Z_{k,k}\right)},
\end{aligned}\end{equation}

\begin{theorem}\label{theorem_1}
For $N \rightarrow \infty$, the value of $\text{SINR}_{k}$ is tightly lower bounded by a deterministic value $\underline{\text{SINR}}_{k}\overset{N\rightarrow \infty}{<}\text{SINR}_{k}$ which is given by
\begin{equation}\begin{aligned}\label{eeq16}
\underline{\text{SINR}}_{k}=\frac{\tau^{2}P_{I}c_{k}^{2}}{\underset{j\neq k}{\sum}c_{k,j}+\tau^{2}P_{I}\bar{c}_{k}^{2}+\bar{c}_{k}^{(1)}+\tilde{c}_{k}+\sigma_{n}^{2}\frac{\tau_{d}+1}{\tau_d}},
\end{aligned}\end{equation}

\noindent
where 
{\fontsize{10}{11}\begin{equation}\begin{aligned}\label{eeq17}\nonumber
&c_{k}=\boldsymbol{p}_{k}^{T}\text{diag}\left(\boldsymbol{\Gamma}_{k}\boldsymbol{C}_{k}\right),~c_{k,j}=\boldsymbol{p}_{j}^{T}\boldsymbol{\Gamma}_{k}\boldsymbol{R}_{j}\boldsymbol{p}_{j},~\tilde{c}_{k}=\boldsymbol{p}^{T}\boldsymbol{\Gamma}_{k}\boldsymbol{R}\;\boldsymbol{p},\\
&\bar{c}_{k}=\boldsymbol{\bar{p}}^{T}\text{diag}\left(\boldsymbol{\Gamma}_{k}\boldsymbol{C}_{k}\right),~\bar{c}_{k}^{(1)}=\boldsymbol{\bar{p}}^{T}\boldsymbol{\Gamma}_{k}\boldsymbol{R}_{k}^{(1)}\boldsymbol{\bar{p}},~\text{and}\\
&\boldsymbol{R}_{k}^{(1)}=\boldsymbol{\bar{R}}_{k}-\tau^{2}P_{I}\boldsymbol{C}_{k}^{2}\boldsymbol{\Gamma}_{k}.
\end{aligned}\end{equation}}
Since $\underline{\text{SINR}}_{k}$ is deterministic (independent of the small-fading randomness, $\mathbb{E}[\underline{\text{SINR}}_{k}]=\underline{\text{SINR}}_{k}$), and based on \eqref{eeq13} and \eqref{eeq16}, $\underline{R}_{k}=\text{log}_{2}(1+\underline{\text{SINR}}_{k})$ is a tight lower bound on the ergodic rate of the attacked user $\text{IU}_{k}$, and known at the CPU. 
\begin{equation}\label{eeq17new}
\underline{R}_{k}=\text{log}_{2}\left(1+\underline{\text{SINR}}_{k}\right)\overset{N\rightarrow \infty}{<}R_{k}
\end{equation}

\end{theorem} 

\begin{IEEEproof}
See Appendix \ref{Appen_A-1}.
\end{IEEEproof}
%
\subsection{Upper Bound on the EH Ergodic Rate}\label{UPEH_rate} 
The received signal at the eavesdropping antenna of the EH in \eqref{eeq5c} can be recast as follows
\begin{equation}\begin{aligned}\label{eeq18}
&y_{E_k}=b_{k} q_{k}+\sum_{j\neq k}b_{j}q_{j}+\hat{b}_{k}z+b+\bar{n},\\
& b_{j}=\boldsymbol{g}_{E}^{T}\boldsymbol{w}_{j},~ \hat{b}_{k}=\boldsymbol{g}_{E}^{T}\boldsymbol{\bar{w}}_{k},~b=\boldsymbol{g}_{E}^{T}\boldsymbol{w}.
\end{aligned}\end{equation}

\noindent
In the following, we assume the worst-case scenario in which the EH has full knowledge of its own channel vectors, $\boldsymbol{g}_{E}$ and $\boldsymbol{g}$; and the beamforming vectors $\{\boldsymbol{w}_{i}\}$. With this worst-case assumption, an upper bound on the ergodic information rate at the EH is given in the following theorem.  
\begin{theorem}\label{theorem_2}
With a worst-case scenario assumption that the EH has full knowledge of its own channel and the beamforming vectors of the IUs, the EH is capable of cancelling the inter-user interference \cite[ Chapter 8]{tse2005fundamentals}. Since the information, $\{q_{i}\}$, the AN signal, $z$, and the energy signal, $\boldsymbol{w}$, are statistically independent, we have the following upper bound, $\overline{R}_{E_k}$, on the ergodic rate of the EH intending to eavesdrop $\text{IU}_{k}$, $R_{E_k}$, given by
\begin{equation}\begin{aligned}\label{eeq13rr2}    
\overline{R}_{E_k}=\text{log}_2\left(1+\mathbb{E}\left[\text{SINR}_{E_k}\right]\right) \geq \\
R_{E_k}=\mathbb{E}\left[\text{log}_2\left(1+\text{SINR}_{E_k}\right)\right],
\end{aligned}\end{equation}

\noindent
for which
\begin{equation}\begin{aligned}\label{eeq20}
&\mathbb{E}\left[\text{SINR}_{E_k}=\frac{\left|b_{k}\right|^2}
  {|\hat{b}_k|^2+|b|^2+\sigma_n^2}\right]\overset{N\to \infty}{\to}\\
  &\frac{\mathbb{E}\left[\left|b_{k}\right|^2\right]}{\mathbb{E}\left[|\hat{b}_k|^2+|b|^2+\sigma_n^2\right]}=\frac{\tau^{2}\phi P_{E}d_{k}^{2}+d_{k}^{(1)}}{\tau^{2}\phi P_{E}\bar{d}_{k}^{2}+\bar{d}_{k}^{(1)}+d+\sigma_{n}^{2}},
\end{aligned}\end{equation}

\noindent 
where

{\fontsize{9.6}{11}\begin{equation}\begin{aligned}\label{eeq17Rep}\nonumber
&d_{k}=\boldsymbol{p}_{k}^{T}\text{diag}\left(\boldsymbol{\Gamma}\boldsymbol{C}_{k}\right),~d_{k}^{(1)}=\boldsymbol{p}_{k}^{T}\boldsymbol{\Gamma}\boldsymbol{R}_{k}^{(2)}\boldsymbol{p}_{k},~d=\boldsymbol{p}^{T}\boldsymbol{\Gamma}\boldsymbol{R}\;\boldsymbol{p},\\
&\bar{d}_{k}=\boldsymbol{\bar{p}}^{T}\text{diag}\left(\boldsymbol{\Gamma}\boldsymbol{C}_{k}\right),~\bar{d}_{k}^{(1)}=\boldsymbol{\bar{p}}^{T}\boldsymbol{\Gamma}\boldsymbol{R}_{k}^{(2)}\boldsymbol{\bar{p}},~ \text{and}\\
&\boldsymbol{R}_{k}^{(2)}=\boldsymbol{\bar{R}}_{k}-\tau^{2}\phi P_{E}\boldsymbol{C}_{k}^{2}\boldsymbol{\Gamma}.
\end{aligned}\end{equation}}

\end{theorem} 

\begin{IEEEproof}
See Appendix \ref{Appen_A-1}. 
\end{IEEEproof} 

Such a worst-case scenario is commonly employed by much of the current research to guarantee maximum information security \cite{wu2016secure,6728676}. Ensuring the confidentiality of the information for the worst-case scenario design ensures confidentiality for more optimistic scenarios. 
\subsection{Lower Bound on the Ergodic Secrecy Rate of $\text{IU}_{k}$}\label{LB_rate} 
Using the lower bound and the upper bound on the information rates at the attacked user $\text{IU}_{k}$ and the EH given in \eqref{eeq17new} and \eqref{eeq13rr2}, we assess the secrecy of information at $\text{IU}_{k}$ in terms of ESR which has the following lower bound     
 \begin{equation}\label{eeq21}
  R_{S_k}\overset{N\rightarrow \infty}{\rightarrow}\left[\underline{R}_{k}-\overline{R}_{E_k}\right]^{+}.
\end{equation} 
%
\subsection{Average Harvested Energy at the EH}\label{Harv_E} 
The EH relies on the dual functionality of its antennas to harvest energy and eavesdrop information simultaneously. The whole signal received via the legitimate antenna is devoted for energy harvesting, while the signal received via the illegitimate antenna is used for information decoding. However, since the CPU blames the EH for the active attack, the received signals via both antennas are accounted for the CPU for energy harvesting. The AHE by the EH intending to eavesdrop $\text{IU}_{k}$ is\footnote{Detailed derivation of the results in \eqref{eeq22} are in Appendix \ref{Appen_A-1}.} 
 {\fontsize{10}{11}\begin{equation}\begin{aligned}\label{eeq22}
 & E_{k}=\zeta\;\mathbb{E}\bigg[\left|b_{k}\right|^{2}+\sum_{j\neq k}\left|b_{j}\right|^{2}+\left|\hat{b}_{k}\right|^{2}+\left|b\right|^{2}+\sum_{j}\left|\tilde{b}_{j}\right|^{2}+\left|\tilde{\hat{b}}_{k}\right|^{2}\\
  &+\left|\tilde{b}\right|^{2}\bigg]=\zeta\Bigg[\tau^{2}\phi P_{E}d_{k}^{2}+d_{k}^{(1)}+\sum_{j\neq k}d_{k,j}+\tau^{2}\phi P_{E}\bar{d}_{k}^{2}+\bar{d}_{k}^{(1)}\\
  &+d+\sum_{j}d_{k,j}+\tilde{d}_{k}+\tau^{2}(1-\phi)P_{E}\tilde{d}^{2}+\tau\sigma_{n}^{2}\tilde{d}^{(1)}\Bigg],
\end{aligned}\end{equation}}
where
 {\fontsize{10}{11}\begin{equation}\begin{aligned}\label{eeq21}\nonumber
  &\tilde{b}_{j}=\boldsymbol{g}^{T}\boldsymbol{w}_{j},~ \tilde{\hat{b}}_{k}=\boldsymbol{g}^{T}\boldsymbol{\bar{w}}_{k},~ \tilde{b}=\boldsymbol{g}^{T}\boldsymbol{w},~ d_{k,j}=\boldsymbol{p}_{j}^{T}\boldsymbol{\Gamma}\boldsymbol{R}_{j}\boldsymbol{p}_{j},\\
  &\tilde{d}_{k}=\boldsymbol{\bar{p}}^{T}\boldsymbol{\Gamma}\boldsymbol{\bar{R}}_{k}\;\boldsymbol{\bar{p}},~\tilde{d}=\boldsymbol{p}^{T}\text{diag}\left(\boldsymbol{\Gamma}\boldsymbol{C}\right),~\text{and}~\tilde{d}^{(1)}=\boldsymbol{p}^{T}\boldsymbol{\Gamma}\boldsymbol{C}^{2}\boldsymbol{p}.
\end{aligned}\end{equation}}

\section{Power Control of Downlink Transmission}\label{Pow_cont} 
\subsection{Problem Formulation}\label{Prob_Formu} 

In our system, a single AP, $\text{AP}_{j}$, transmits a set of $M+2$ different types of signals, $\{\{[\boldsymbol{w}_{i}]_{j}q_{i}\}_{i},~[\boldsymbol{\bar{w}}_{k}]_{j}z,~[\boldsymbol{w}]_{j}\}$. With the random geometric distribution of the APs with respect to the IUs and the EH, the power control in the cell-free MIMO system has an advantage over the conventional MIMO that different users have different subsets of dominant serving APs. In the long-term, the CPU can achieve a fair and secured SWIPT transmission towards the IUs and the EH by balancing the average levels of transmit powers at the APs within the power limits of each AP. The power control aims to maximize the worst-case ESR, $\min_{k} R_{S_k}$, with a constraint on the minimum AHE requirement of the legitimate EH. Therefore, our constrained problem is 
\begin{subequations}\label{CFeeq25}
\begin{flalign}
& \underset{\left\{\boldsymbol{p}_{i}\right\},~\boldsymbol{\bar{p}},~\boldsymbol{p}}{\text{maximize}}\hspace{.8cm} \min_{k} R_{S_k}\nonumber\\
&\text{subject to}\nonumber \\
& E_{k}\ge \bar{E},~ \forall k,\label{CFeeq25a}\\
&\left[\sum_{i=1}^{M}\mathbb{E}[\boldsymbol{w}_{i}\boldsymbol{w}_{i}^{H}]+\mathbb{E}[\boldsymbol{\bar{w}}_{k}\boldsymbol{\bar{w}}_{k}^{H}]+\mathbb{E}[\boldsymbol{w}\boldsymbol{w}^{H}]\right]_{j,j}\leq P_{t},~ \forall j,~ \forall k,
\label{CFeeq25b}
\end{flalign}
\end{subequations}  

\noindent
where $P_{t}$ is the available power budget at each AP. The constraint \eqref{CFeeq25b} guarantees the average power consumption at each AP is within the limit, $P_{t}$. Problem \eqref{CFeeq25} is non-convex since the objective function is a logarithm of multiplicative fractional functions. Without loss of generality, we assume that \eqref{CFeeq25} is always feasible and focus on solving it. We use the exponential variable substitution method used in \cite{alageli2017optimization} and \cite{alageli2017optimal} to transform the logarithmic objective function of \eqref{CFeeq25} into an equivalent linear function. By using the properties of logarithmic and exponential functions, the objective function of \eqref{CFeeq25} can be expressed as $\text{log}_{e}2~\text{ln}(e^{u_{k}-s_{k}}e^{v_{k}-t_{k}})$ where 
\begin{subequations}\begin{flalign}
&e^{u_{k}}=\tau^{2}P_{I}c_{k}^{2}+\underset{j\neq k}{\sum}c_{k,j}+\tau^{2}P_{I}\bar{c}_{k}^{2}+\bar{c}_{k}^{(1)}+\tilde{c}_{k}+\sigma_{n}^{2}\frac{\tau_{d}+1}{\tau_d}\label{CFeeq26a}\\
&e^{s_{k}}=\underset{j\neq k}{\sum}c_{k,j}+\tau^{2}P_{I}\bar{c}_{k}^{2}+\bar{c}_{k}^{(1)}+\tilde{c}_{k}+\sigma_{n}^{2}\frac{\tau_{d}+1}{\tau_d}\label{CFeeq26b}\\
&e^{t_{k}}=\tau^{2}\phi P_{E}d_{k}^{2}+d_{k}^{(1)}+\tau^{2}\phi P_{E}\bar{d}_{k}^{2}+\bar{d}_{k}^{(1)}+d+\sigma_{n}^{2}\label{CFeeq26c}\\
&e^{v_{k}}=\tau^{2}\phi P_{E}\bar{d}_{k}^{2}+\bar{d}_{k}^{(1)}+d+\sigma_{n}^{2}.\label{CFeeq26d}
\end{flalign}\end{subequations}  

Since the logarithmic functions are monotonically increasing in their arguments, then \eqref{CFeeq25} can be recast as 
\begin{subequations}\label{CFeeq27}
\begin{flalign}
& \underset{\substack{\left\{\boldsymbol{p}_{i}\right\},~\boldsymbol{\bar{p}},~\boldsymbol{p}\\\{u_{k},~s_{k},~t_{k},~v_{k}\}}}{\text{maximize}}\hspace{.8cm} \min_{k} \left(u_{k}-s_{k}+v_{k}-t_{k}\right)\nonumber\\
&\text{subject to}\nonumber \\
&\tau^{2}P_{I}c_{k}^{2}+\underset{j\neq k}{\sum}c_{k,j}+\tau^{2}P_{I}\bar{c}_{k}^{2}+\bar{c}_{k}^{(1)}+\tilde{c}_{k}+\sigma_{n}^{2}\frac{\tau_{d}+1}{\tau_d} \nonumber\\
&\geq e^{u_{k}},~\forall~k,\label{CFeeq27a}\\
&\underset{j\neq k}{\sum}c_{k,j}+\tau^{2}P_{I}\bar{c}_{k}^{2}+\bar{c}_{k}^{(1)}+\tilde{c}_{k}+\sigma_{n}^{2}\frac{\tau_{d}+1}{\tau_d},\nonumber\\
&\leq e^{\bar{s}_{k}}\left(s_{k}-\bar{s}_{k}+1\right),~\forall~k,\label{CFeeq27b}\\
&\tau^{2}\phi P_{E}d_{k}^{2}+d_{k}^{(1)}+\tau^{2}\phi P_{E}\bar{d}_{k}^{2}+\bar{d}_{k}^{(1)}+d+\sigma_{n}^{2}\nonumber\\
&\leq e^{\bar{t}_{k}}\left(t_{k}-\bar{t}_{k}+1\right),~\forall~k,\label{CFeeq27c}\\
&\tau^{2}\phi P_{E}\bar{d}_{k}^{2}+\bar{d}_{k}^{(1)}+d+\sigma_{n}^{2}\geq e^{v_{k}},~\forall~k,\\
&\eqref{CFeeq25a},~\eqref{CFeeq25b}.
\label{CFeeq27d}
\end{flalign}
\end{subequations}

Our new objective in \eqref{CFeeq27} is monotonically increasing with $\min_{k} R_{S_k}$. The constraints \eqref{CFeeq27a}--\eqref{CFeeq27d} bound the slack variables $u_{k},~s_{k},~t_{k},~v_{k}$ of the objective function within their limits defined in \eqref{CFeeq26a}--\eqref{CFeeq26d}. The exponential variables $e^{s_{k}}$ and $e^{t_{k}}$ are linearized as $e^{\bar{s}_{k}}(s_{k}-\bar{s}_{k}+1)$ and $e^{\bar{t}_{k}}(t_{k}-\bar{t}_{k}+1)$. $\bar{s}_{k},~\bar{t}_{k}$ are the initial values around which $e^{s_{k}}$ and $e^{t_{k}}$ are linearized. 

The formulation in \eqref{CFeeq27} is still non-convex since the right-hand sides of the constraints \eqref{CFeeq27a}--\eqref{CFeeq27d} contain expressions which are nonlinear in the optimization variables (the power control factors $\{\{\boldsymbol{p}_{i}\},~\boldsymbol{\bar{p}},~\boldsymbol{p}\}$), such as $c_{k}^{2}=\left(\boldsymbol{p}_{k}^{T}\text{diag}\left(\boldsymbol{\Gamma}_{k}\boldsymbol{C}_{k}\right)\right)^{2}$. These nonlinearities arise from the per AP per user power control (specific for cell-free massive MIMO systems) where each AP has its own transmit power constraint. In comparison, these nonlinearities do not exist in the power control for the conventional (collocated) massive MIMO systems in which the constraint is on the total transmit power from all collocated antennas \cite{alageli2018swipt}. To deal with these nonlinearities, we introduce a new method of cooperative balancing of the transmit powers at the APs via relaxed SDP formulation which has been proved to be optimal as will be described in the next subsection. 
 
\subsection{SDP Formulation for Optimal Power Control}\label{Prob_Formu} 
In this subsection, we reformulate the non-convex problem \eqref{CFeeq27} into a relaxed SDP convex problem. To achieve this, the nonlinear expressions in the power control factors $\{\{\boldsymbol{p}_{i}\},~\boldsymbol{\bar{p}},~\boldsymbol{p}\}$ are represented as linear expressions in terms of new rank-one positive semidefinite matrix variables $\{\{\boldsymbol{P}_{i}=\boldsymbol{p}_{i}\;\boldsymbol{p}_{i}^{H}\},~\boldsymbol{\bar{P}}=\boldsymbol{\bar{p}}\;\boldsymbol{\bar{p}}^{H},~\boldsymbol{P}=\boldsymbol{p}\;\boldsymbol{p}^{H}\}$. For instance, given that $k$ is the index of the IU under attack, the expression of $c_{k}^{2}$ can be recast in an SDP form as
\begin{equation}\begin{aligned}\label{CFeeq29}
c_{k}^{2}&=\left(\boldsymbol{p}_{k}^{T}\text{diag}\left(\boldsymbol{\Gamma}_{k}\boldsymbol{C}_{k}\right)\right)^{2}=\boldsymbol{p}_{k}^{T}\text{diag}\left(\boldsymbol{\Gamma}_{k}\boldsymbol{C}_{k}\right)\text{diag}\left(\boldsymbol{\Gamma}_{k}\boldsymbol{C}_{k}\right)^{T}\boldsymbol{p}_{k}\\
&=\text{tr}\left(\boldsymbol{p}_{k}\boldsymbol{p}_{k}^{T}\text{diag}\left(\boldsymbol{\Gamma}_{k}\boldsymbol{C}_{k}\right)\text{diag}\left(\boldsymbol{\Gamma}_{k}\boldsymbol{C}_{k}\right)^{T}\right)=\text{tr}\left(\boldsymbol{P}_{k}\boldsymbol{A}_{k}\right),
\end{aligned}\end{equation}

\noindent   
where ${\fontsize{9.6}{11}\boldsymbol{A}_{k}=\text{diag}(\boldsymbol{\Gamma}_{k}\boldsymbol{C}_{k})~\text{diag}(\boldsymbol{\Gamma}_{k}\boldsymbol{C}_{k})^{T}}$. In a comparable way, the rest of the expressions $\{c_{k,j},~\bar{c}_{k}^{2},~\bar{c}_{k}^{(1)},~\tilde{c}_{k}\}$, $\{d_{k}^{2},~d_{k}^{(1)},~d,~\bar{d}_{k}^{2},~\bar{d}_{k}^{(1)}\}$ and $\{d_{k,j},~\tilde{d}_{k},~\tilde{d}^{2},~\tilde{d}^{(1)}\}$ in \eqref{CFeeq27} can be transformed into linear expressions in terms of $\{\{\boldsymbol{P}_{i}\},~\boldsymbol{\bar{P}},~\boldsymbol{P}\}$. With these transformations, we can recast the non-convex problem in \eqref{CFeeq27} into a convex relaxed\footnote{The formulation in \eqref{CFeeq30} does not impose any constraints on the rank of $\{\{\boldsymbol{P}_{i}\},~\boldsymbol{\bar{P}},~\boldsymbol{P}\}$, i.e, $\{\text{rank}(\boldsymbol{P}_{i})\},~\text{rank}(\boldsymbol{\bar{P}}),~\text{rank}(\boldsymbol{P})\leq N$.} SDP formulation as in \eqref{CFeeq30} at the top of the next page, where $\mathbb{S}=\{\{\boldsymbol{P}_{i}\},~\boldsymbol{\bar{P}},~\boldsymbol{P},~\{u_{k},~s_{k},~t_{k},~v_{k}\}\}$ is the set of optimization variables and
\begin{figure*}[!h]
{\fontsize{10}{11}
\begin{subequations}\label{CFeeq30}
\begin{flalign}
& \underset{\mathbb{S}}{\text{maximize}}\hspace{.8cm} \min_{k} \left(u_{k}-s_{k}+v_{k}-t_{k}\right)\nonumber\\
&\text{subject to}\nonumber\\
&\hspace{0cm}\tau^{2}P_{I}\text{tr}\left(\boldsymbol{P}_{k}\boldsymbol{A}_{k}\right)+\underset{j\neq k}{\sum}\text{tr}\left(\boldsymbol{P}_{j}\boldsymbol{A}_{k,j}\right)+\tau^{2}P_{I}\text{tr}\left(\boldsymbol{\bar{P}}\boldsymbol{A}_{k}\right) +\text{tr}\left(\boldsymbol{\bar{P}}\boldsymbol{\bar{A}}_{k}\right)+\text{tr}\left(\boldsymbol{P}\boldsymbol{\tilde{A}}_{k}\right)+\sigma_{n}^{2}\frac{\tau_{d}+1}{\tau_d}\geq e^{u_{k}},~\forall~k,\label{CFeeq30a}\\
&\hspace{0cm}\underset{j\neq k}{\sum}\text{tr}\left(\boldsymbol{P}_{j}\boldsymbol{A}_{k,j}\right)+\tau^{2}P_{I}\text{tr}\left(\boldsymbol{\bar{P}}\boldsymbol{A}_{k}\right)+\text{tr}\left(\boldsymbol{\bar{P}}\boldsymbol{\bar{A}}_{k}\right)+\text{tr}\left(\boldsymbol{P}\boldsymbol{\tilde{A}}_{k}\right)+\sigma_{n}^{2}\frac{\tau_{d}+1}{\tau_d}\leq e^{\bar{s}_{k}}\left(s_{k}-\bar{s}_{k}+1\right),~\forall~k,\label{CFeeq30b}\\
&\hspace{0cm}\tau^{2}\phi P_{E}\text{tr}\left(\boldsymbol{P}_{k}\boldsymbol{B}_{k}\right)+\text{tr}\left(\boldsymbol{P}_{k}\boldsymbol{\bar{B}}_{k}\right)+\tau^{2}\phi P_{E}\text{tr}\left(\boldsymbol{\bar{P}}\boldsymbol{B}_{k}\right)+\text{tr}\left(\boldsymbol{\bar{P}}\boldsymbol{\bar{B}}_{k}\right)+\text{tr}\left(\boldsymbol{P}\boldsymbol{B}\right)+\sigma_{n}^{2}\leq e^{\bar{t}_{k}}\left(t_{k}-\bar{t}_{k}+1\right),~\forall~k,\label{CFeeq30c}\\
&\tau^{2}\phi P_{E}\text{tr}\left(\boldsymbol{\bar{P}}\boldsymbol{B}_{k}\right)+\text{tr}\left(\boldsymbol{\bar{P}}\boldsymbol{\bar{B}}_{k}\right)+\text{tr}\left(\boldsymbol{P}\boldsymbol{B}\right)+\sigma_{n}^{2}\geq e^{v_{k}},~\forall~k,\label{CFeeq30d}\\
&\zeta\Bigg(\tau^{2}\phi P_{E}\text{tr}\left(\boldsymbol{P}_{k}\boldsymbol{B}_{k}\right)+\text{tr}\left(\boldsymbol{P}_{k}\boldsymbol{\bar{B}}_{k}\right)+\sum_{j\neq k}\text{tr}\left(\boldsymbol{P}_{j}\boldsymbol{\tilde{B}}_{j}\right)+\tau^{2}\phi P_{E}\text{tr}\left(\boldsymbol{\bar{P}}\boldsymbol{B}_{k}\right)+\text{tr}\left(\boldsymbol{\bar{P}}\boldsymbol{\bar{B}}_{k}\right)+\text{tr}\left(\boldsymbol{P}\boldsymbol{B}\right)+\sum_{j}\text{tr}\left(\boldsymbol{P}_{j}\boldsymbol{\tilde{B}}_{j}\right)\nonumber\\
&~~~~+\text{tr}\left(\boldsymbol{\bar{P}}\boldsymbol{\tilde{B}}_{k}\right)+\tau^{2}(1-\phi) P_{E}\text{tr}\left(\boldsymbol{P}\boldsymbol{\ddot{B}}\right)+\tau\sigma_{n}^{2}\text{tr}\left(\boldsymbol{P}\boldsymbol{\hat{B}}\right)\Bigg)\geq \bar{E},~\forall~k,\label{CFeeq30e}\\
&\text{tr}\left(\boldsymbol{P}_{k}\boldsymbol{D}_{l}\boldsymbol{\bar{R}}_{k}\right)+\sum_{j\neq k}\text{tr}\left(\boldsymbol{P}_{j}\boldsymbol{D}_{l}\boldsymbol{R}_{j}\right)+\text{tr}\left(\boldsymbol{\bar{P}}\boldsymbol{D}_{l}\boldsymbol{\bar{R}}_{k}\right)+\text{tr}\left(\boldsymbol{P}\boldsymbol{D}_{l}\boldsymbol{R}\right)-P_{t}\leq 0,~\forall~l,~\forall~k,\label{CFeeq30f}\\
&\left\{\boldsymbol{P}_{k}\right\},~\boldsymbol{\bar{P}},~\boldsymbol{P}\succeq 0. \label{CFeeq30g}
\end{flalign}
\end{subequations}}
\hrulefill
\end{figure*} 

\noindent
{\fontsize{10}{11}\begin{equation}\begin{aligned}\label{eeq17Rep2}\nonumber
&\boldsymbol{A}_{k,j}=\boldsymbol{\Gamma}_{k}\boldsymbol{R}_{j},~\boldsymbol{\bar{A}}_{k}=\boldsymbol{\Gamma}_{k}\boldsymbol{R}_{k}^{(1)},~\boldsymbol{\tilde{A}}_{k}=\boldsymbol{\Gamma}_{k}\boldsymbol{R}\\
&\boldsymbol{B}_{k}=\text{diag}\left(\boldsymbol{\Gamma}\boldsymbol{C}_{k}\right)\text{diag}\left(\boldsymbol{\Gamma}\boldsymbol{C}_{k}\right)^{H},~\boldsymbol{\bar{B}}_{k}=\boldsymbol{\Gamma}\boldsymbol{R}_{k}^{(2)},~\boldsymbol{B}=\boldsymbol{\Gamma}\boldsymbol{R},\\
&\boldsymbol{\tilde{B}}=\boldsymbol{\Gamma}\boldsymbol{R}_{j},~\boldsymbol{\ddot{B}}=\text{diag}\left(\boldsymbol{\Gamma}\boldsymbol{C}\right)\text{diag}\left(\boldsymbol{\Gamma}\boldsymbol{C}\right)^{H},~\boldsymbol{\hat{B}}=\boldsymbol{\Gamma}\boldsymbol{C}^{2},~\text{and}\\
&\boldsymbol{\tilde{B}}_{k}=\boldsymbol{\Gamma}\boldsymbol{\bar{R}}_{k}.
\end{aligned}\end{equation}}
The constraints \eqref{CFeeq30e} and \eqref{CFeeq30f} are an SDP recast of \eqref{CFeeq25a} and \eqref{CFeeq25b}, respectively. The constraint \eqref{CFeeq30f} is equivalent to \eqref{CFeeq25b}, where $\boldsymbol{D}_{l}\in\mathcal{R}^{N\times N}$ has zero entries except $[\boldsymbol{D}_{l}]_{l,l}=1$. This equivalent representation in \eqref{CFeeq30f} is required to facilitate the proof of Theorem \ref{theorem_3} presented in Appendix \ref{Appen_A-1}. 

 The formulation in \eqref{CFeeq30} is convex and can be solved iteratively based on the initial value update method given in Algorithm 1. It can be shown that the complex-valued SDP problem \eqref{CFeeq31} (which is equivalent to \eqref{CFeeq30}) contains: $M+2$ semidefinite complex-valued $N\times N$ matrix variables, $5M+1$ real scalar variables, $6M+NM+2$ constraints on matrix variable of size $N\times N$, and $M$ constraints on scalar variables. The complexity (in terms of number of complex operations) of obtaining a per iteration solution of \eqref{CFeeq30} within accuracy $\epsilon$ is asymptotically upper bounded by $\mathcal{O}(M^{4}N^{\frac{9}{2}}\log(\frac{1}{\epsilon}))$ \cite{5447068}. This result assumes unstructured input data matrices. However, the optimization solver (such as SeDuMi employed by CVX software \cite{grant2010cvx}) can exploit the structure of input data matrices -- for example, the structure of single non-zero element matrices $\{\boldsymbol{D}_{l}\}$ -- to reduce the computational complexity \cite{5447068}.   
  \begin{algorithm}[t]
   \algsetup{linenosize=\tiny}
  \footnotesize
 \caption{Algorithm for solving problem (\ref{CFeeq30})}
 \begin{algorithmic}[1]
 \STATE Initialize $\{\bar{s}_{k}^{[n]}\}$ and $\{\bar{t}_{k}^{[n]}\}$, $n=1$.
 \STATE \textbf{Repeat}
 \STATE Solve problem (\ref{CFeeq30}) and calculate $\{s_{k}^{[n]}\}$ and $\{t_{k}^{[n]}\}$.
 \STATE Increment $n=n+1$.
 \STATE Update the initial values $\bar{s}_{k}^{[n]}=\text{ln}(e^{\bar{s}_{k}^{[n-1]}}(s_{k}^{[n-1]}-\bar{s}_{k}^{[n-1]}+1))$ and $\bar{t}_{k}^{[n]}=\text{ln}(e^{\bar{t}_{k}^{[n-1]}}(t_{k}^{[n-1]}-\bar{t}_{k}^{[n-1]}+1))$.
 \STATE \textbf{Until}\; \text{Convergence}.
 \end{algorithmic}
 \end{algorithm}
%
\subsection{Global Optimality of the SDP Formulation}\label{Glob_Opt} 
To investigate the optimality of the solution obtained by \eqref{CFeeq30}, let us rewrite \eqref{CFeeq30} in the equivalent form in \eqref{CFeeq31} by replacing the objective $\min_{k} \left(u_{k}-s_{k}+v_{k}-t_{k}\right)$ by a new slack viable $\pi$ and $K$ linear constraints as  
\begin{subequations}\label{CFeeq31}
\begin{flalign}
 &\underset{\substack{\left\{\boldsymbol{P}_{i}\right\},~\boldsymbol{\bar{P}},~\boldsymbol{P}\\\{\text{diag}([u_{k},~s_{k},~t_{k},~v_{k}])\},~\pi}}{\text{maximize}}\hspace{.8cm} \pi \nonumber\\
&\text{subject to}~~~~~ \text{diag}([u_{k},~s_{k},~t_{k},~v_{k}])-\pi\boldsymbol{I_4}\succeq 0,~ \forall k,\label{CFeeq31a}\\
&\hspace{2cm}\eqref{CFeeq30a}\text{--}\eqref{CFeeq30g}.
\label{CFeeq31b}
\end{flalign}
\end{subequations} 

\noindent
By examining \eqref{CFeeq31} with the first-order and the second-order conditions of convexity, we have
\begin{equation}\label{CFeeq32}
\frac{\partial \pi }{\partial \pi}=1,~~\text{and}~~\frac{\partial^{2} \pi }{\partial \pi^{2}}=0.
\end{equation}

\noindent
This means that \eqref{CFeeq30} is convex with an affine objective function. Since the constraints of \eqref{CFeeq30} are differentiable and there are no constraints on the domain of the optimization variables $\{\boldsymbol{P}_{i}\},~\boldsymbol{\bar{P}},~\boldsymbol{P}\in \mathcal{S}^{+}$, $\{u_{k},~s_{k},~t_{k},~v_{k}\},~\pi\in \mathcal{R}$, then Slater's condition holds and the solution obtained by solving \eqref{CFeeq30} is globally optimal subject to: 1) satisfying the rank requirement of $\{\boldsymbol{P}_{i}\},~\boldsymbol{\bar{P}}~~\text{and}~~\boldsymbol{P}$; 2) and the convergence of the constraints \eqref{CFeeq30b} and \eqref{CFeeq30c} (which results in the convergence of the iterative problem \eqref{CFeeq30}). 

\subsubsection{Rank-one Optimality}

Generally, the optimality of the solutions obtained via SDP programming might require a rank higher than one. The rank requirement for the optimality of the solutions obtained by SDP problems has been investigated in \cite[Lemma 3.1]{5233822} which is quoted as:

\begin{lemma}\label{Lemma_2} 
\textit{Suppose that the separable SDP (P1) and its dual (D1) are solvable. Then, problem (P1) has always an optimal solution $\{\boldsymbol{X}_{1}^{\star},~\dots,~\boldsymbol{X}_{L}^{\star}\}$ such that
\begin{equation}\label{CFeeq33}
\sum_{l=1}^L \text{rank}^2\left(\boldsymbol{X}_{l}^{\star}\right)\leq M.
\end{equation}}
\end{lemma}

\noindent
where $\{\boldsymbol{X}_{1},~\dots,~\boldsymbol{X}_{L}\}$ are the semi-definite matrix variables of (P1), $\{\boldsymbol{X}_{1}^{\star},~\dots,~\boldsymbol{X}_{L}^{\star}\}$ are their optimal values and $M$ (for
\eqref{CFeeq33} only) is the number of constraints. Nevertheless, for our problem \eqref{CFeeq30}, the obtained solution $\{\boldsymbol{P}_{i}^{\star}\},~\boldsymbol{\bar{P}}^{\star}$ and $\boldsymbol{P^{\star}}$ needs to satisfy the rank-one structure $\{\text{rank}(\boldsymbol{P}_{i}^{\star})\},~\text{rank}(\boldsymbol{\bar{P}}^{\star}),~\text{rank}(\boldsymbol{P}^{\star})=1$ which complies with the optimality condition \eqref{CFeeq33} in Lemma \ref{Lemma_2}. The compliance of $\{\boldsymbol{P}_{i}^{\star}\},~\boldsymbol{\bar{P}}^{\star}$ and $\boldsymbol{P^{\star}}$ with rank-one requirement is given in the following theorem

\begin{theorem}\label{theorem_3}
Given that $\mathbb{S}^{\star}=\{\{\boldsymbol{P}_{i}^{\star}\},~\boldsymbol{\bar{P}}^{\star},~\boldsymbol{P^{\star}},~\{u_{k}^{\star},~s_{k}^{\star},~t_{k}^{\star},$ $v_{k}^{\star}\}\}$ is the solution obtained by solving \eqref{CFeeq30}, then, the optimized power control factor matrices $\{\boldsymbol{P}_{i}^{\star}\},~\boldsymbol{\bar{P}}^{\star},~\boldsymbol{P^{\star}}$ always
satisfy the rank-one constraint, i.e., $\{\text{rank}(\boldsymbol{P}_{i}^{\star})\},~\text{rank}(\boldsymbol{\bar{P}}^{\star}),~\text{rank}(\boldsymbol{P}^{\star})=1$. 
\end{theorem}   

\begin{IEEEproof}
See Appendix \ref{Appen_A0}. 
\end{IEEEproof}
%
\subsubsection{Convergence of the Iterative Problem} \label{converg}
Here, we prove that the iterative optimization \eqref{CFeeq30} converges to a globally optimal value, and the objective value (which is monotonically increasing with $\min_{k} R_{S_k}$) is increasing with the iterations. To facilitate our proof, let us introduce the following results:

\begin{lemma}
\label{lemma_3}
For arbitrary real values of $x$ and $\bar{x}\neq x$, the first order approximation $e^{\bar{x}}\left(x-\bar{x}+1\right)$ is always an underestimate of $e^{x}$, i.e.
\begin{equation} \label{CFeeq34}
e^{\bar{x}}\left(x-\bar{x}+1\right)\leq e^{x},~\forall~\bar{x}<x,~\bar{x}>x.  
\end{equation}
\end{lemma}

\begin{IEEEproof}
See Appendix \ref{Appen_C}.
\end{IEEEproof} 

\begin{lemma}
\label{lemma_4}
For arbitrary real values of $x$ and $\bar{x}\neq x$, the successive first order approximations of $e^{x}$; $f^{[n]}=e^{\bar{x}^{[n]}}\left(x-\bar{x}^{[n]}+1\right)$ and $f^{[n+1]}=e^{\bar{x}^{[n+1]}}\left(x-\bar{x}^{[n+1]}+1\right)$, $e^{\bar{x}^{[n+1]}}=f{[n]}$; always satisfy $f^{[n+1]}>f^{[n]}$ for $\bar{x}^{[n]}\neq x$.  
\end{lemma}

\begin{IEEEproof}
See Appendix \ref{Appen_C}. 
\end{IEEEproof}

Without loss of generality, we assume that \eqref{CFeeq30} is feasible in its first iteration. Since our problem is convex and Slater's condition holds (see \eqref{CFeeq32} and the paragraph that follows), constraints \eqref{CFeeq30b} and \eqref{CFeeq30c} can strictly hold. With the first order linearisation in \eqref{CFeeq30b} and \eqref{CFeeq30c}, and according to Lemma \ref{lemma_3}, the constraints \eqref{CFeeq30b} and \eqref{CFeeq30c} are tighter than their original formulations in \eqref{CFeeq27b} and \eqref{CFeeq27c}, i.e., the feasibility region of \eqref{CFeeq27} is smaller than and a subregion of the feasibility region of \eqref{CFeeq30}. Therefore, any non-converged solution is suboptimal.

According to Lemma \ref{lemma_4}, and since the constraints \eqref{CFeeq30b} and \eqref{CFeeq30c} are initialized in the $n\text{th}$ iteration by the optimal values obtained at the $(n-1)\text{th}$ preceding iteration such as $e^{\bar{s}_{k}^{[n]}}=e^{\bar{s}_{k}^{[n-1]}}(s_{k}^{\star^{[n-1]}}-\bar{s}_{k}^{[n-1]}+1)$ and $e^{\bar{t}_{k}^{[n]}}=e^{\bar{t}_{k}^{[n-1]}}(t_{k}^{\star^{[n-1]}}-\bar{t}_{k}^{[n-1]}+1)$, $\forall~k$, the feasibility of the $(n-1)\text{th}$ iteration will ensure the feasibility of the succeeding $n\text{th}$ iteration. Furthermore, the feasibility region at the $n\text{th}$ iteration is larger than the feasibility region at the $(n-1)\text{th}$ iteration and contains it. This ensures that the optimized objective value is monotonically increasing with the successive iteration. Given that the constrained values in \eqref{CFeeq30b} and \eqref{CFeeq30c} are finitely bounded (since both constraints are linear and the power budget at every AP is finite, $\leq P_{t}$), therefore, it can be concluded that the increasing optimized objective value will certainly converge, let us say at the $n$th iteration, i.e.
\begin{equation} \label{CFeeq36}
\begin{aligned}
e^{\bar{s}_{k}^{{[n]}}}\left(s_{k}^{\star^{[n]}}-\bar{s}_{k}^{{[n]}}+1\right)=e^{\bar{s}_{k}^{{[n-1]}}}\left(s_{k}^{\star^{[n-1]}}-\bar{s}_{k}^{{[n-1]}}+1\right).
\end{aligned}
\end{equation}

\noindent
By solving the updating method, $e^{\bar{s}_{k}^{[n]}}=e^{\bar{s}_{k}^{[n-1]}}(s_{k}^{\star^{[n-1]}}-\bar{s}_{k}^{[n-1]}+1)$ and \eqref{CFeeq36}, we have $s_{k}^{\star^{[n]}}=\bar{s}_{k}^{{[n]}}$, and $e^{\bar{s}_{k}^{{[n]}}}(s_{k}^{\star^{[n]}}-\bar{s}_{k}^{{[n]}}+1)=e^{s_{k}^{\star^{[n]}}}$. This indicates that the constraint \eqref{CFeeq30b} converges to its original nonlinearized form. Likewise, the constraint \eqref{CFeeq30c} converges to its original nonlinearized form.

\section{Evaluations}\label{Eval} 

In this section, we evaluate the asymptotic performance of our SWIPT cell-free massive MIMO system. The APs are randomly located on a two dimensional Euclidean area $A_{a}$ based on an homogeneous PPP $\Phi_a$ with an intensity $\lambda_a$. The IUs and the EH are randomly located on a two dimensional Euclidean area $A_{u}$ with the origins of $A_{a}$ and $A_{u}$ coincident (please refer to footnote 1 regarding this assumption). The large-scale fading coefficients $\{\gamma_{i,j},~\gamma_{j}\}$ are modeled with the standard distance-based model as $\gamma_{i,j}\triangleq d_{i,j}^{-\alpha}~10^{\frac{\nu_{i,j}}{10}}$ and $\gamma_{j}\triangleq d_{j}^{-\alpha}~10^{\frac{\nu}{10}}$, where $d_{i,j}$ and $d_{j}$ are the distances from $\text{IU}_i$ and the EH to $\text{AP}_j$, respectively. $\alpha$ is the pathloss exponent and $\nu,~\nu_{i,j}\sim\mathcal{CN}(0,\sigma^{2})$ are the shadow fading coefficients with standard division $\sigma$. All users experience independent shadow fading, i.e., $\nu_{i,j}$ and $\nu_{i,j}$s are independent random variables (RVs). $P$, $P_{E}$ and $P_{t}$ are the training power budget at every IU, the training power budget at the EH, and the transmit power budget at every AP, respectively. $\tau$ and $\zeta$ are the length of the training sequences and the energy harvesting efficiency at the EH, respectively. Unless otherwise stated, and for referencing convenience, the selected values of system parameters are listed in Table II.     
 \begin{center}
 \begin{table}[t]\label{tab:table1}  
\centering\footnotesize
\renewcommand{\arraystretch}{1.3}
\captionsetup{justification=centering, labelsep=newline}
\caption {\scshape Selected Values of System Parameters}
    \begin{tabular}{ |m{1.3cm}|m{3.5cm}| }
    \hline
    Parameter & Value   \\ \hline
    $A_{a}$, $\lambda_a$& $1\times 1$ $\text{Km}^{2}$, $1.5\times 10^{-4}~m^{-1}$\\ \hline
     $A_{u}$, $M$& $300\times 300$ $\text{m}^{2}$, 3  \\ \hline
     $\alpha$, $\sigma$& 2.5, $8~\text{dB}$ \cite{rappaport1996wireless} \\ \hline
    $P$, $P_{E}$, $P_{t}$& $1~W$, $1~W$, $500~mW$\\ \hline
    $\tau$, $\tau_d$, $\zeta$& 10, 10, 0.5 \cite{le2008efficient}\\ \hline
    \end{tabular}
\end{table}
\end{center}

 \begin{figure}[t]
\begin{center}
\captionsetup{justification=centering}
\includegraphics[width=.4\textwidth,trim = 1cm 0cm 1cm 1cm, clip]{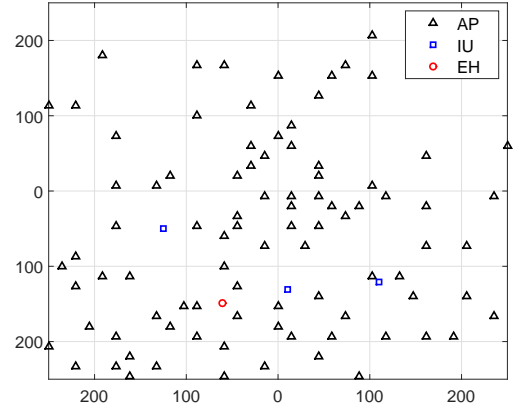}
\caption{AP-user deployment, $N=145$ and $M=3$.}
\label{fig1}
\end{center}
\end{figure}

\begin{figure}[h]
\begin{center}
\captionsetup{justification=centering}
\includegraphics[width=.4\textwidth,trim = 1cm 0cm 1cm 1cm, clip]{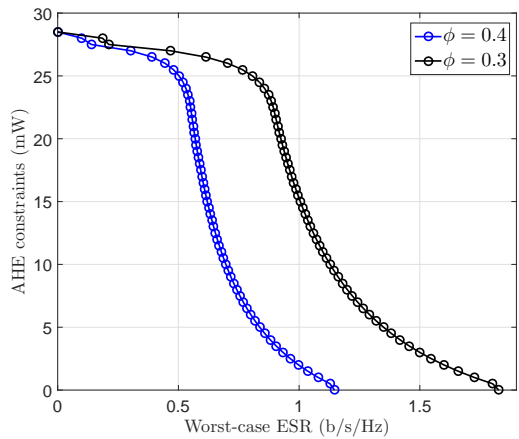}
\caption{E-R regions of colocated MIMO.}
\label{fig3c}
\end{center}
\end{figure}

\begin{figure}[h]
\begin{center}
\captionsetup{justification=centering}
\includegraphics[width=.4\textwidth,trim = 1cm 0cm 1cm 1cm, clip]{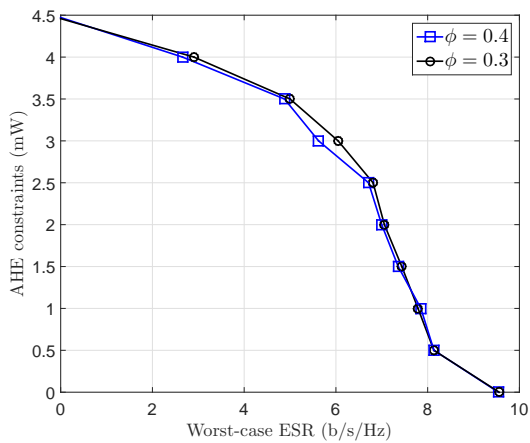}
\caption{E-R regions of cell-free MIMO.}
\label{fig4c}
\end{center}
\end{figure}

\begin{figure}[h]
\begin{center}
\captionsetup{justification=centering}
\includegraphics[width=.4\textwidth,trim = 1cm 0cm 1cm 1cm, clip]{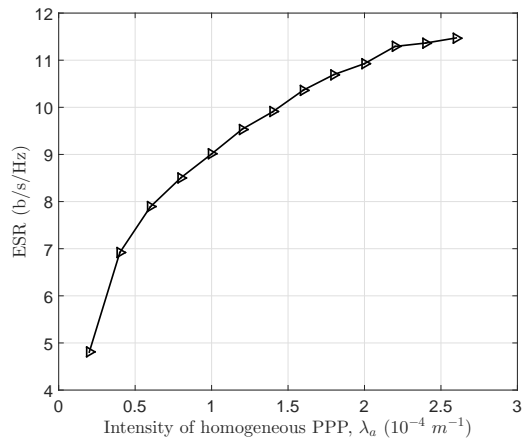}
\caption{ESR versus the intensity of homogeneous PPP, $\lambda_a$.} 
\label{fig5c}
\end{center}
\end{figure}

Fig. \ref{fig1} shows the AP-user deployment geometry of a realization in which the number of APs is $N=145$ ($\mathbb{E}(N)=\lambda_a A_{a}=150$), $M=3$ IUs and one EH zoomed into the central $500\times 500$ $\text{m}^{2}$. 

The SWIPT secrecy performance is presented by the E-R plot which relates the achievable worst-case ESR, $\min_{k} R_{S_k}$, to the constraint on the minimum AHE by the EH, $\bar{E}$. The larger the area under the E-R curve, the better the SWIPT performance. Our design analyses are made based on the asymptotic assumption $N\to\infty$, then, the system's performance can be examined for a realistic scenario of a large but finite number of APs. 

In colocated MIMO systems, the user exhibits a constant average path-loss to all of the base station's (BS's) colocated antennas, and that average path-loss varies from one user to another based on the user's location. In contrast, in cell-free MIMO systems, the average path-loss of a given user varies from one AP to another. Intuitively, this property of randomly distributed APs is anticipated to increase the efficiency of power control in tackling the active eavesdropping. For fair comparisons between the performance of cell-free and colocated MIMO systems, a comparable model of single-cell colocated massive MIMO system is derived such that: 1) the total number of colocated antennas at the BS is equal to the total number of APs, $N$; 2) the average value of a user's pathloss to the BS in colocated MIMO (all users experience equal pathlosses) is equal to the average value of the users' pathlosses in cell-free MIMO; 3) the total transmit power is equal for both system, and the power limits at the colocated MIMO is per antenna; 4) the antennas of the BS are uncorrelated. Defining $\bar{\gamma}_{i}$ and $\bar{\gamma}$ as the pathlosses of $\text{IU}_i$ and the EH in the colocated MIMO system, respectively, we have $\bar{\gamma}_{i}=\sum_{j}\frac{\gamma_{i,j}}{N}$ and $\bar{\gamma}=\sum_{j}\frac{\gamma_{j}}{N}$. The downlink beamfoming and power control of the colocated MIMO can be performed by the same methods used for cell-free MIMO. 

Fig. \ref{fig3c} shows the E-R regions of the colocated MIMO system for two different values of active eavesdropping power corresponding to training power splitting factors $\phi=0.3$ and $\phi=0.4$. It can be noticed that there is a tradoff between the ESR and the constraint on the AHE. As the AHE constraint increases, more downlink transmission resources are optimized to satisfy the increase in AHE constraint at the expense of the ESR which tends to decrease. Also, there is a clear gap between the ESR performances at different levels of active eavesdropping powers. The larger the eavesdropping power the lower the ESR.

Fig. \ref{fig4c} shows the E-R regions of the cell-free MIMO system for the same values of active eavesdropping powers used for colocated MIMO system. By comparing Fig. \ref{fig3c} and Fig. \ref{fig4c}, it can be noticed that the cell-free MIMO outperforms the colocated MIMO within the interval in which the harvested energy constraint is low and vice versa. The cell-free MIMO is also found to be more immune to the increase in the active eavesdropping power than the colocated MIMO. In colocated MIMO, all antennas contribute to the AHE by an equal average value which is not the case for the cell-free MIMO. Therefore, the colocated MIMO is more efficient at power transfer than the cell-free MIMO. The difference between channel gains of the IU and the EH in the cell-free system offers the optimizer more freedom to balance the tradeoff between the information, AN and the energy signal powers than in the case of the colocated MIMO system. That justifies the advantage of cell-free MIMO over the colocated MIMO in the feasible region (the low AHE constraint region).  

Fig. \ref{fig5c} shows how the density of APs affects the secrecy performance. The achievable worst-case ESR is measured versus a set of practically large values of AP densities $\lambda_a = 10^{ -4}\times[0.2,~0.4,~ ...,~2.6]~m^{-1}$. The values of the worst-case ESR in Fig. \ref{fig5c} are obtained by Monte Carlo simulation averaged over 50 independent realizations of AP deployments, with $\bar{E}=0$ and $\phi=0.3$. As expected, as the
AP density (which is directly proportional to $\mathbb{E}[N]$) increases, the worst-case ESR
increases.
  
\begin{figure}[t]
\begin{center}
\captionsetup{justification=centering}
\includegraphics[width=.4\textwidth,trim = 1cm 0cm 1cm 1cm, clip]{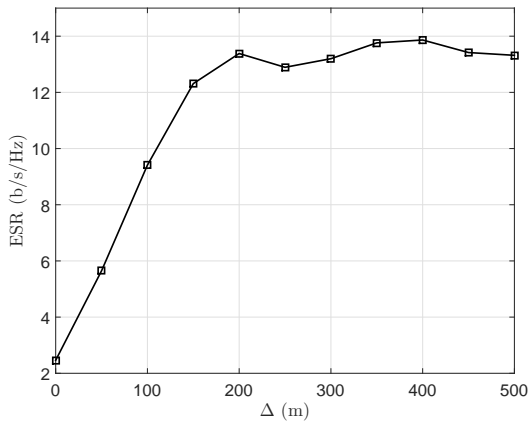}
\caption{ESR versus the separation between the IU and the EH for $\bar{E}=0$, $\phi=0.5$ and  and $\alpha = 2$.}
\label{fig3}
\end{center}
\end{figure}

\begin{figure}[t]
\begin{center}
\captionsetup{justification=centering}
\includegraphics[width=.4\textwidth,trim = 0cm 0cm 1cm 1cm, clip]{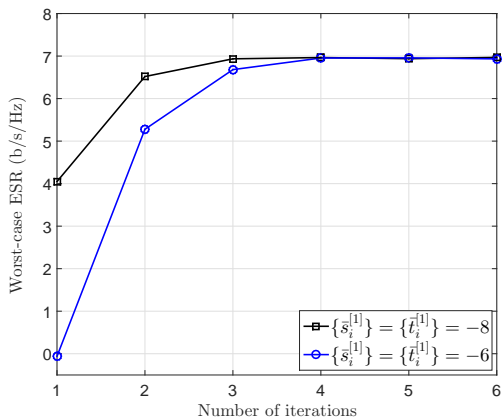} 
\caption{Convergence speed of the iterative program in \eqref{CFeeq30} for $\bar{E}=2~mW$ and $\phi=0.4$.}
\label{fig4}
\end{center}
\end{figure}

The secrecy performance is affected by the relative location of the attacked IU with respect to the EH. Fig. \ref{fig3} shows the ESR performance for the case where the system comprises one IU and one EH lying on the $x$-axis symmetrically around the origin of the APs' deployment given in Fig \ref{fig1}. The results represent the achieved ESR for different separation distances between the IU and the EH, $\Delta=[0,~100,~\dots,500]$. As the separation $\Delta$ increases, the ESR performance improves. This can be justified since as the separation increases, the APs subsets that dominantly serve the IU and the EH become more distinctive. Beyond a certain value of $\Delta>200$, the achieved ESR starts to saturate since the dominant subsets of the APs that serve the IU and the EH remain unchanged, but the position of each user within its set varies. The value $\Delta=0$ means that the IU and the EH are colocated, i.e., $\boldsymbol{\Gamma}_{1}=\boldsymbol{\Gamma}$.

Fig. \ref{fig4} shows the convergence speed of Algorithm 1 at $\phi=0.4$, $\bar{E}=2~mW$ and the initial values are selected arbitrarily as $\bar{s}_{i}^{[1]}=\bar{t}_{i}^{[1]}=[-8,~-6]~\forall~i$. As discussed in Subsection \ref{converg}, the optimized objective value is increasing with iterations until convergence.

\section{Conclusions}\label{Sec_Conc}   
In this paper, relaxed SDP programming has been proposed to optimize a nonlinear power control of the downlink transmission in a SWIPT cell-free massive MIMO system in the presence of an information-untrusted dual-antenna active EH. The downlink SWIPT transmissions include: information, AN and energy signals beamformed towards the IUs, legitimate and illegitimate antennas of the EH, respectively. Analytic expressions for the AHE and a tight lower bound on the ESR were derived with taking into account the IUs' knowledge attained by downlink effective precoded-channel training. It has been proved that the proposed SDP iterative problem can always achieve a converged rank-one globally optimal solution. A fair comparison between the proposed cell-free and the colocated massive MIMO systems showed that the cell-free MIMO outperformed the colocated MIMO over the interval in which the AHE constraint is low and vice versa. Also, cell-free MIMO was more immune to the increase in the active eavesdropping power than the colocated MIMO.  

\appendices 
\section{Proofs of Lemma \ref{lemma1c}, Theorems \ref{theorem_1} and \ref{theorem_2}}  \label{Appen_A-1} 
\subsection{Proof of Lemma \ref{lemma1c}}
Since the spectral radius of the diagonal matrices $\boldsymbol{\Gamma}_{i}$, $\boldsymbol{\Gamma}$ and $\sqrt{\boldsymbol{\Gamma}_{i}\boldsymbol{\Gamma}}$ are bounded \cite[Lemma 2]{alageli2018swipt}, then by expanding $\boldsymbol{y}_{i}^{H}\boldsymbol{y}_{i}$ followed by applying Corollary 1 in \cite{evans2000large} we get 
\begin{equation}\label{ApA200}
\boldsymbol{y}_{i}^{H}\boldsymbol{y}_{i}-\tau^{2}P_{I}\text{tr}\left(\boldsymbol{\Gamma}_{i}\right)-N\tau\sigma_{n}^{2}\stackrel{N\to\infty}{\to}\delta_{ij}~\tau^{2}\phi P_{E}\text{tr}\left(\boldsymbol{\Gamma}\right),
\end{equation}

\noindent
which satisfies the asymptotic convergence in \eqref{eeq3}. This concludes the proof.

\subsection{Proof of Theorem \ref{theorem_1}} \label{Theo1}
Before commencing our proof, let us introduce the following result.

\begin{lemma}\label{lemma_6}
For a non-negative bounded RV $X_{1}\leq U$, $U$ is a positive real value, and a symmetrical zero mean RV $X_{2}$, the non-negative RV $Y=X_{1}+X_{2}$ is upper bounded as $Y\leq 2U$
\end{lemma}

\begin{IEEEproof} 
We have $P((X_{1}+X_{2})<0)=0$, then $P(U+X_{2}<0)=0$, i.e., $P(X_{2}<-U)=0$. By symmetry of distribution, $P(X_{2}>U)=0$ which implies $X_{2}\leq U$. Therefore, $Y\leq2U$. This concludes the proof.  
\end{IEEEproof}
Let $\text{IU}_k$ be the attacked IU. Based on \eqref{eeq2}, \eqref{eeq4a}, \eqref{eeq6a} and \eqref{eeq8} we have
\begin{equation}\begin{aligned}\label{ApA21}
&\left|\hat{a}_{k,k}\right|^{2}=\tau^{2}P_{I}\left|\boldsymbol{\bar{h}}_{k}^{H}\boldsymbol{\Gamma}_{k}^{\frac{1}{2}}\text{diag}\left(\boldsymbol{p}_{k}\right)\boldsymbol{C}_{k}\boldsymbol{\Gamma}_{k}^{\frac{1}{2}}\boldsymbol{\bar{h}}_{k}\right|^{2}\\
&\hspace{1.3cm}+\left|\boldsymbol{\bar{h}}_{k}^{H}\boldsymbol{\Gamma}_{k}^{\frac{1}{2}}\text{diag}\left(\boldsymbol{p}_{k}\right)\boldsymbol{C}_{k}\boldsymbol{\tilde{h}}_{k}\right|^{2}+\left|\frac{\boldsymbol{n}_{k}\boldsymbol{\psi}_{d_k}^{*}}{\tau_d}\right|^{2}\\
\end{aligned}\end{equation}

\noindent
where $\boldsymbol{\tilde{h}}_{k}=\boldsymbol{\hat{h}}_{k}-\tau\sqrt{P_I}\boldsymbol{C}_{k}\boldsymbol{h}_{k}=\tau \sqrt{\phi P_{E}}\;\boldsymbol{g}_{E} + \boldsymbol{N}\boldsymbol{\psi}_{k}^{*}$, and the entries of $\boldsymbol{\tilde{h}}_{k}$, $\boldsymbol{h}_{k}$ and $\boldsymbol{n}_{k}$ are statistically independent. Using Corollary 1 in \cite{evans2000large} and Lemma 2 in \cite{alageli2018swipt}, the first term in \eqref{ApA21} asymptotically converges to the deterministic value 
\begin{equation}\begin{aligned}\label{ApA22}
&\Delta_{1}=\tau^{2}P_{I}\text{tr}^{2}(\boldsymbol{\Gamma}_{k}\text{diag}(\boldsymbol{p}_{k})\boldsymbol{C}_{k})\\
&=\tau^{2}P_{I}\left(\boldsymbol{p}_{k}^{T}\text{diag}(\boldsymbol{\Gamma}_{k}\boldsymbol{C}_{k})\right)^{2}=\tau^{2}P_{I}\text{tr}\left(\boldsymbol{P}_{k}\boldsymbol{A}_{k}\right)=\tau^{2}P_{I}c_{k}^{2}=\\
&\tau^{2}P_{I}\bigg(\sum_{j}\gamma_{k,j}^{2}\bar{c}_{k,j}^{2}p_{k,j}+\sum_{\mathbb{I}}\gamma_{k,j}\gamma_{k,m}\bar{c}_{k,j}\bar{c}_{k,m}\sqrt{p_{k,j}p_{k,m}}\bigg),
\end{aligned}\end{equation} 

\noindent
where $\mathbb{I}=\{\{k,j\}_{j}\times\{k,m\}_{m}|\{k,j\}\neq\{k,m\}\}$ and $\bar{c}_{k,j}=[\boldsymbol{C}_{k}]_{j,j}$. With the assumption that the noise variance $\sigma_{n}^{2}\ll \tau\phi P_{E}$, we can approximate the sum of the second and the third terms in \eqref{ApA21} as $\Delta_{2}=\tau^{2} \phi P_{E}|\boldsymbol{\bar{h}}_{k}^{H}\boldsymbol{\Gamma}_{k}^{\frac{1}{2}}\text{diag}\left(\boldsymbol{p}_{k}\right)\boldsymbol{C}_{k}\boldsymbol{g}_{E}|^{2}$, which is equivalent to 
\begin{equation}\begin{aligned}\label{ApA23}
\Delta_{2}=&\tau^{2}\phi P_{E}\bigg(\sum_{j=1}^{N}\gamma_{k,j}\gamma_{j}\bar{c}_{k,j}^{2}p_{k,j}~\kappa_{j}+\\
&\sum_{\mathbb{I}}\sqrt{\gamma_{k,j}\gamma_{j}\gamma_{k,m}\gamma_{m}}\bar{c}_{k,j}\bar{c}_{k,m}\sqrt{p_{k,j}p_{k,m}}~\kappa_{j,m}\bigg),
\end{aligned}\end{equation}

\noindent
where $\kappa_{j}=|\hat{h}_{k,j}\hat{g}_{E_j}|^{2}$ (equivalent to the product of two independent exponential RVs of parameter 1) is a non-negative RV with the mean value $\mathbb{E}[\kappa_{j}]=1$. $\kappa_{j,m}=\hat{h}_{k,j}\hat{g}_{E_j}\hat{h}_{k,m}\hat{g}_{E_m}$, $j\neq m$, is a zero mean RV with a symmetric distribution \cite{hamedani1985product,chen1983sum}. Since $\Delta_{2}$ is always positive, i.e., $P(\Delta_{2}<0)=0$, then, by applying Theorem A in \cite{berend2010improved} (which defines an upper bound on the sum of non-negative RVs) and Lemma \ref{lemma_6}, $\Delta_{2}$ is upper bounded by a deterministic value as   
\begin{equation}\begin{aligned}\label{ApA24}
\Delta_{2}\leq \overline{\Delta}_{2}=4 e\tau^{2}\phi P_{E}\sum_{j=1}^{N}\gamma_{k,j}\gamma_{j}\bar{c}_{k,j}^{2}p_{k,j}.
\end{aligned}\end{equation}

Given that the additive terms that constitute $\Delta_{1}$ in \eqref{ApA22} and the upper bound of $\Delta_{2}$, $\overline{\Delta}_{2}$, in \eqref{ApA24} are of a finite order of magnitude, then,
asymptotically, we have $\Delta_{1}\overset{N\to \infty}{\to}\mathcal{O}\left(N^{2}\right)$ and $\overline{\Delta}_{2}\overset{N\to \infty}{\to}\mathcal{O}\left(N\right)$. Therefore, as $N\to \infty$, $\Delta_{1}$ and $\Delta_{2}$ differ by $\mathcal{O}\left(N\right)$ order of magnitude which implies that the bound $\left|\hat{a}_{k,k}\right|^{2}\geq \Delta_{1}=\tau^{2}P_{I}\text{tr}\left(\boldsymbol{P}_{k}\boldsymbol{A}_{k}\right)$ is tight. Based on this result, \eqref{eeq14}, \eqref{eeq16}, and since $\text{SINR}_{k}$ and $\underline{\text{SINR}}_{k}$ share the same denominator, then $\text{SINR}_{k}>\underline{\text{SINR}}_{k}$ in \eqref{eeq16} is of the same degree of tightness. To validate the tightness of $\text{SINR}_{k}\overset{N\rightarrow \infty}{>}\underline{\text{SINR}}_{k}$ numerically, Table III presents the values of $\Delta_{1}$, $\overline{\Delta}_{2}$ and $\frac{\Delta_{1}}{\overline{\Delta}_{2}}$ for different realizations of $\{\boldsymbol{\Gamma}_{i}\}$ and $\boldsymbol{\Gamma}$ at a large average value of $N=100$, and at $\bar{E}=5~mW$. The optimized values of $\{\boldsymbol{P}_{i}\}$, $\boldsymbol{\bar{P}}$ and $\boldsymbol{P}$ used to obtain the values of $\Delta_{1}$ are used to calculate corresponding value of $\overline{\Delta}_{2}$. The obtained results validate our analysis.
\begin{center}
\begin{table}[t]\label{tab:table1}
\centering\footnotesize
\caption { \scshape Relative values of $\Delta_{1}$ and $\overline{\Delta}_{2}$} 
    \begin{tabular}{ |p{1.2 cm}| p{1.27 cm} | p{1.27 cm}| p{1.27 cm} |p{1.27 cm} | }
    \hline
    Realization & 1st & 2nd & 3rd & 4th  \\ \hline
    $\Delta_{1}$                 & $9.8\times 10^{-3}$  & $7.3\times 10^{-3}$& $1.6\times 10^{-2}$& $4.6\times 10^{-3}$\\ \hline
    $\overline{\Delta}_{2}$           & $6.9\times 10^{-5}$& $1.1\times 10^{-4}$& $3.8\times 10^{-5}$& $1.2\times 10^{-4}$\\ \hline
    $\frac{\Delta_{1}}{\overline{\Delta}_{2}}$&$1.4\times 10^{2}$ & $0.65\times 10^{2}$ & $4.1\times 10^{2}$& $0.3\times 10^{2}$ \\ \hline
    \end{tabular}
\end{table}
\end{center}    
The values of $\text{var}(Z_{k,k})=\sum_{j\neq k} c_{k,j}+\tau^{2}P_{I}\bar{c}_{k}^{2}+\bar{c}_{k}^{(1)}+\tilde{c}_{k}$ and $\text{var}(\tilde{a}_{k,k})=\sigma_{n}^{2}\frac{\tau_{d}+1}{\tau_d}$ (as in \eqref{eeq14}-\eqref{eeq16}) can be calculated as follows. We have 
\begin{equation}\begin{aligned}\label{ApffAc1}
&\tilde{a}_{k,k}=a_{k,k}-\mathbb{E}\left[a_{k,k}|\hat{a}_{k,k}\right]=a_{k,k}-\hat{a}_{k,k}=a_{k,k}-\frac{y_{I_k}}{\tau_d}\\
&=a_{k,k}-\frac{\boldsymbol{y}_{I_k}\psi_{d_k}^{*}}{\tau_d}=a_{k,k}-\frac{\left(\sum_{j=1}^{M}a_{k,j}\boldsymbol{\psi}_{d_j}^{T}+\boldsymbol{n}_{k}\right)\psi_{d_k}^{*}}{\tau_d}\\
&=a_{k,k}-\frac{a_{k,k}\tau_{d}+\boldsymbol{n}_{k}\psi_{d_k}^{*}}{\tau_d}=\frac{\boldsymbol{n}_{k}\psi_{d_k}^{*}}{\tau_d}.
\end{aligned}\end{equation}
\noindent
The second equality follows from \eqref{eeqjk12}. The third and fifth equalities follow from substituting \eqref{eeq8} and \eqref{eeq6a}, respectively. Since $\mathbb{E}(\boldsymbol{n}_{k})=0$, then $\mathbb{E}(\tilde{a}_{k,k})=0$ and therefore
\begin{equation}\begin{aligned}\label{Apdse1}
\text{var}\left(\tilde{a}_{k,k}\right)=\mathbb{E}\left[|\tilde{a}_{k,k}|^{2}\right]=\mathbb{E}\left[\frac{\boldsymbol{n}_{k}\psi_{d_k}^{*}\psi_{d_k}^{T}\boldsymbol{n}_{k}^{H}}{\tau_{d}^{2}}\right]=\frac{\tau_{d}\sigma_{n}^{2}}{\tau_{d}^{2}}=\frac{\sigma_{n}^{2}}{\tau_{d}}.
\end{aligned}\end{equation}
\noindent
$\text{var}\left(\tilde{a}_{k,k}\right)+\sigma_{n}=\frac{\sigma_{n}}{\tau_{d}^{2}}+\sigma_{n}^{2}=\sigma_{n}^{2}\frac{\tau_{d}+1}{\tau_{d}}$ is equivalent to the last term in the denominator of \eqref{eeq16}. Regarding $Z_{k,k}$, the additive terms that constitute $Z_{k,k}$ in \eqref{eeq11} are zero mean statistically independent RVs since the entries of $\{a_{k,j}\}_{j\neq k}$, $\boldsymbol{\bar{w}}_{k}$, $\boldsymbol{w}$ and $n_{i}$ are zero mean statistically independent RVs. Therefore,
\begin{equation}\begin{aligned}\label{Aprte1}
&\text{var}(Z_{k,k})=\\
&\sum_{j\neq k} \mathbb{E}\left[|a_{k,j}|^{2}\right]+\mathbb{E}\left[|\boldsymbol{h}_{k}^{T}\boldsymbol{\bar{w}}_{k}|^{2}\right]+\mathbb{E}\left[|\boldsymbol{h}_{k}^{T}\boldsymbol{w}|^{2}\right]+\mathbb{E}\left[|n_{i}|^{2}\right].
\end{aligned}\end{equation}
\noindent
The expectations in \eqref{Aprte1} are calculated as follows.
\begin{equation}\begin{aligned}\label{Aprffss1}
&\mathbb{E}\left[|a_{k,j}|^{2}\right]=\mathbb{E}\left[|\boldsymbol{h}_{k}^{T}\boldsymbol{w}_{j}|^{2}\right]=\mathbb{E}\left[|\boldsymbol{h}_{k}^{T}\boldsymbol{w}_{j}\boldsymbol{w}_{j}^{H}\boldsymbol{h}_{k}^{*}|^{2}\right]\\
&=\mathbb{E}\left[|\boldsymbol{\bar{h}}_{k}^{T}\boldsymbol{\Gamma}_{k}^{\frac{1}{2}}\text{diag}\left(\boldsymbol{p}_{j}\right)\boldsymbol{\hat{h}}_{j}^{*}\boldsymbol{\hat{h}}_{j}^{T}\text{diag}\left(\boldsymbol{p}_{j}\right)\boldsymbol{\Gamma}_{k}^{\frac{1}{2}}\boldsymbol{\bar{h}}_{k}^{*}|^{2}\right]\\
&=\mathbb{E}\left[|\boldsymbol{\bar{h}}_{k}^{T}\boldsymbol{\Gamma}_{k}^{\frac{1}{2}}\text{diag}\left(\boldsymbol{p}_{j}\right)\mathbb{E}\left[\boldsymbol{\hat{h}}_{j}^{*}\boldsymbol{\hat{h}}_{j}^{T}\right]\text{diag}\left(\boldsymbol{p}_{j}\right)\boldsymbol{\Gamma}_{k}^{\frac{1}{2}}\boldsymbol{\bar{h}}_{k}^{*}|^{2}\right]\\
&=\text{tr}\left(\boldsymbol{\Gamma}_{k}^{\frac{1}{2}}\text{diag}\left(\boldsymbol{p}_{j}\right)\boldsymbol{R}_{j}\text{diag}\left(\boldsymbol{p}_{j}\right)\boldsymbol{\Gamma}_{k}^{\frac{1}{2}}\right)=\boldsymbol{p}_{j}^{T}\boldsymbol{\Gamma}_{k}\boldsymbol{R}_{j}\boldsymbol{p}_{j}\\
&=\text{tr}\left(\boldsymbol{P}_{j}\boldsymbol{\Gamma}_{k}\boldsymbol{R}_{j}\right)=\text{tr}\left(\boldsymbol{P}_{j}\boldsymbol{A}_{k,j}\right)=c_{k,j}.
\end{aligned}\end{equation}
\noindent
The third equality in \eqref{Aprffss1} is obtained by substituting the values of $\boldsymbol{h}_{k}$ and $\boldsymbol{w}_{j}$ from \eqref{eeq4a}. In the fourth equality, the expectation is moved to $\boldsymbol{\hat{h}}_{j}^{*}\boldsymbol{\hat{h}}_{j}^{T}$ based on the statistical independence between $\boldsymbol{\bar{h}}_{k}$ and $\boldsymbol{\hat{h}}_{j}$. The fifth equality follows since the entries of $\boldsymbol{\bar{h}}_{k}$ are zero mean unit variance independent RVs. The sixth and the seventh equalities follow since the matrices $\boldsymbol{\Gamma}_{k}$, $\text{diag}\left(\boldsymbol{p}_{j}\right)$ and $\boldsymbol{R}_{j}$ are diagonal. The forms $\boldsymbol{p}_{j}^{T}\boldsymbol{\Gamma}_{k}\boldsymbol{R}_{j}\boldsymbol{p}_{j}$ and $\text{tr}(\boldsymbol{P}_{j}\boldsymbol{\Gamma}_{k}\boldsymbol{R}_{j})$ are identical to those used in \eqref{eeq16} and \eqref{CFeeq30a}-\eqref{CFeeq30b}, respectively. For $\mathbb{E}[|\boldsymbol{h}_{k}^{T}\boldsymbol{\bar{w}}_{k}|^{2}]$, we substitute the value of $\boldsymbol{\bar{w}}_{k}$ from \eqref{eeq4b}, \eqref{eeq2a} and \eqref{eeq2c} as
\begin{equation}\begin{aligned}\label{Aprss1}
&\mathbb{E}\left[|\boldsymbol{h}_{k}^{T}\boldsymbol{\bar{w}}_{k}|^{2}\right]=\mathbb{E}\left[|\boldsymbol{h}_{k}^{T}\text{diag}\left(\boldsymbol{\bar{p}}\right)\boldsymbol{C}_{k}\left(\tau\sqrt{P_{I}}\boldsymbol{h}_{k}^{*}+\boldsymbol{\tilde{h}}_{k}^{*}\right)|^{2}\right]\\
&=\tau^{2}P_{I}\mathbb{E}\left[|\boldsymbol{h}_{k}^{T}\text{diag}\left(\boldsymbol{\bar{p}}\right)\boldsymbol{C}_{k}\boldsymbol{h}_{k}^{*}|^{2}\right]+\mathbb{E}\left[|\boldsymbol{h}_{k}^{T}\text{diag}\left(\boldsymbol{\bar{p}}\right)\boldsymbol{C}_{k}\boldsymbol{\tilde{h}}_{k}^{*}|^{2}\right]\\
&=\tau^{2}P_{I}|\text{tr}\left(\text{diag}\left(\boldsymbol{\bar{p}}\right)\boldsymbol{C}_{k}\boldsymbol{\Gamma}_{k}\right)|^{2}\\
&\hspace{.5cm}+\mathbb{E}\left[\boldsymbol{\bar{h}}_{k}^{T}\boldsymbol{\Gamma}_{k}^{\frac{1}{2}}\text{diag}\left(\boldsymbol{\bar{p}}\right)\boldsymbol{C}_{k}\mathbb{E}\left[\boldsymbol{\tilde{h}}_{k}^{*}\boldsymbol{\tilde{h}}_{k}^{T}\right]\boldsymbol{C}_{k}\text{diag}\left(\boldsymbol{\bar{p}}\right)\boldsymbol{\Gamma}_{k}^{\frac{1}{2}}\boldsymbol{\bar{h}}_{k}^{*}\right]\\
&=\tau^{2}P_{I}|\boldsymbol{\bar{p}}^{T}\text{diag}\left(\boldsymbol{C}_{k}\boldsymbol{\Gamma}_{k}\right)|^{2}+\boldsymbol{\bar{p}}^{T}\boldsymbol{\Gamma}_{k}\boldsymbol{R}_{k}^{(1)}\boldsymbol{\bar{p}}\\
&=\tau^{2}P_{I}\bar{c}_{k}^{2}+\bar{c}_{k}^{(1)}=\tau^{2}P_{I}\text{tr}\left(\boldsymbol{\bar{P}}\boldsymbol{A}_{k}\right)+\text{tr}\left(\boldsymbol{\bar{P}}\boldsymbol{\bar{A}}_{k}\right).
\end{aligned}\end{equation}
\noindent
where $\boldsymbol{\tilde{h}}_{k}=\delta_{ik}\;\tau \sqrt{\phi P_{E}}\;\boldsymbol{g}_{E} + \boldsymbol{N}\boldsymbol{\psi}_{i}^{*}$. The second equality follows from the statistical independent between $\boldsymbol{h}_{k}$ and $\boldsymbol{\tilde{h}}_{k}$. The first term after the third equality follows from applying Corollary 1 in \cite{evans2000large}. In the second term in the third equality, the expectation is moved to $\boldsymbol{\tilde{h}}_{k}^{*}\boldsymbol{\tilde{h}}_{k}^{T}$ based on the statistical independence between $\boldsymbol{\bar{h}}_{k}$ and $\boldsymbol{\tilde{h}}_{k}$. The first term in the fourth equality follows since the matrices $\boldsymbol{\Gamma}_{k}$, $\boldsymbol{C}_{k}$ and $\text{diag}\left(\boldsymbol{\bar{p}}\right)$ are diagonal. The second term in the fourth equality follows since the entries of $\boldsymbol{\bar{h}}_{k}$ are zero mean unit variance independent RVs and the matrices $\boldsymbol{\Gamma}_{k}$ and $\boldsymbol{R}_{k}^{(1)}$ are diagonal. The form which is after the fourth equality is identical to that used in \eqref{eeq16}, while the SDP form which is after the sixth equality is identical to that used in \eqref{CFeeq30a}-\eqref{CFeeq30b}. For $\mathbb{E}[|\boldsymbol{h}_{k}^{T}\boldsymbol{w}|^{2}]$, we substitute the value of $\boldsymbol{w}$ from \eqref{eeq4c} as 
\begin{equation}\begin{aligned}\label{Aprss1}
&\mathbb{E}\left[|\boldsymbol{h}_{k}^{T}\boldsymbol{w}|^{2}\right]=\mathbb{E}\left[|\boldsymbol{h}_{k}^{T}\text{diag}\left(\boldsymbol{p}\right)\boldsymbol{\hat{g}}^{*}|^{2}\right]\\
&=\mathbb{E}\left[\boldsymbol{\bar{h}}_{k}^{T}\boldsymbol{\Gamma}_{k}^{\frac{1}{2}}\text{diag}\left(\boldsymbol{p}\right)\mathbb{E}\left[\boldsymbol{\hat{g}}^{*}\boldsymbol{\hat{g}}^{T}\right]\text{diag}\left(\boldsymbol{p}\right)\boldsymbol{\Gamma}_{k}^{\frac{1}{2}}\boldsymbol{\bar{h}}_{k}^{*}\right]\\
&=\boldsymbol{p}^{T}\boldsymbol{\Gamma}_{k}\boldsymbol{R}~\boldsymbol{p}=\tilde{c}_{k}=\text{tr}\left(\boldsymbol{P}\boldsymbol{\tilde{A}}_{k}\right).
\end{aligned}\end{equation} 
\noindent
In the second equality, the expectation is moved to $\boldsymbol{\hat{g}}^{*}\boldsymbol{\hat{g}}^{T}$ based on the statistical independence between $\boldsymbol{\bar{h}}_{k}$ and $\boldsymbol{\hat{g}}$. The third equality follows since the entries of $\boldsymbol{\bar{h}}_{k}$ are zero mean unit variance independent RVs and the matrices $\boldsymbol{\Gamma}_{k}$ and $\boldsymbol{R}$ are diagonal. The form which is after the third equality is identical to that used in \eqref{eeq16}, while the SDP form which is after the fifth equality is identical to that used in \eqref{CFeeq30a}-\eqref{CFeeq30b}.

\subsection{Proof of Theorem \ref{theorem_2}}
Based on the assumption that the EH has a full knowledge of the IUs' beamforming vectors and its own channel, the EH is capable of cancelling the inter-user interference \cite{tse2005fundamentals}. Furthermore, the information, AN and energy signals; $\{q_{i}\}$, $z$ and $\boldsymbol{w}$; are statistically independent. Therefore, based on the concavity of the logarithmic function, applying Jensen's inequality (which has been proven to be tight and suitable for characterizing the performance of massive MIMO systems \cite{7727938}) will result in the following upper bound on the ergodic rate at the EH  
\begin{equation}\begin{aligned}\label{ApAc1}
  \overline{R}_{E_k}=\text{log}_2\left(1+\mathbb{E}\left[\text{SINR}_{E_k}\right]\right)>\mathbb{E}\left[\text{log}_2\left(1+\text{SINR}_{E_k}\right)\right],
\end{aligned}\end{equation}
\noindent
where $\text{SINR}_{E_k}$ is the $\text{SINR}$ at the EH when attacking $\text{IU}_{k}$. As defined in \eqref{eeq20}, ${\fontsize{9.5}{11}\text{SINR}_{E_k}}$ ${\fontsize{9.5}{11}=\frac{X_{k}}{Y_{k}}}$, ${\fontsize{9.5}{11}X_{k}=\mathbb{E}[|\boldsymbol{g}_{E}^{T}\boldsymbol{w}_{k}q_{k}|^{2}]=|b_{k}|^2}$ and ${\fontsize{9.5}{11}Y_{k}}$ ${\fontsize{9.5}{11}=\mathbb{E}[|\boldsymbol{g}_{E}^{T}(\boldsymbol{\bar{w}}_{k}z+\boldsymbol{w})+\bar{n}|^{2}]}$ ${\fontsize{9.5}{11}=|\hat{b}_k|^2+|b|^2+\sigma_n^2}$ . Using the multivariate Taylor expansion, 
${\fontsize{9.5}{11}\mathbb{E}[\text{SINR}_{E_k}]}$ can be expanded as \cite{van2000mean}
\begin{equation}\begin{aligned}\label{ApAc2}
  &\mathbb{E}\left[\text{SINR}_{E_k}\right]=\mathbb{E}\left[\frac{X_k}{Y_k}\right]\\
  &=\frac{\mathbb{E}[X_k]}{\mathbb{E}[Y_k]}-\frac{\text{cov}(X_k,Y_k)}{(\mathbb{E}[Y_k])^2}+
  \frac{\text{var}(Y_k)}{(\mathbb{E}[Y_k])^2}~\frac{\mathbb{E}[X_k]}{\mathbb{E}[Y_k]}+R.
\end{aligned}\end{equation}
\noindent
where $R = f(\text{var}(Y_k), \text{cov}(X_k, Y_k))$ is the reminder of
the series expansion. We have 

\begin{equation}\begin{aligned}\label{ApAc3}
&\mathbb{E}[X_k]=\mathbb{E}[|b_{k}|^2]=\mathbb{E}[|\boldsymbol{g}_{E}^{T}\boldsymbol{w}_{k}|^{2}]\\
&=\mathbb{E}\left[\left|\boldsymbol{g}_{E}^{T}\text{diag}\left(\boldsymbol{p}_{k}\right)\left(\tau\sqrt{\phi P_E}\boldsymbol{C}_{k}\boldsymbol{g}_{E}^{*}+\boldsymbol{\tilde{h}}_{k}^{(2)}\right)\right|^{2}\right]=\tau^{2}\phi P_{E}\\
&\mathbb{E}\left[\left|\boldsymbol{\bar{g}}_{E}^{T}\boldsymbol{\Gamma}^{\frac{1}{2}}\text{diag}\left(\boldsymbol{p}_{k}\right)\boldsymbol{C}_{k}\boldsymbol{\Gamma}^{\frac{1}{2}}\boldsymbol{\bar{g}}_{E}^{*}\right|^{2}\right]+\mathbb{E}\left[\left|\boldsymbol{g}_{E}^{T}\text{diag}\left(\boldsymbol{p}_{k}\right)\boldsymbol{\tilde{h}}_{k}^{(2)}\right|^{2}\right],
\end{aligned}\end{equation}  
\noindent
where   
\begin{equation}\begin{aligned}
&\mathbb{E}\left[\left|\boldsymbol{\bar{g}}_{E}^{T}\boldsymbol{\Gamma}^{\frac{1}{2}}\text{diag}\left(\boldsymbol{p}_{k}\right)\boldsymbol{C}_{k}\boldsymbol{\Gamma}^{\frac{1}{2}}\boldsymbol{\bar{g}}_{E}^{*}\right|^{2}\right]=\left|\boldsymbol{p}_{k}^{T}\text{diag}\left(\boldsymbol{\Gamma}\boldsymbol{C}_{k}\right)\right|^{2}\\
&=\text{tr}\left(\boldsymbol{P}_{k}\text{diag}\left(\boldsymbol{\Gamma}\boldsymbol{C}_{k}\right)\text{diag}\left(\boldsymbol{\Gamma}\boldsymbol{C}_{k}\right)^{H}\right)=\text{tr}\left(\boldsymbol{P}_{k}\boldsymbol{B}_{k}\right),\label{ApAc4a}\\
\end{aligned}\end{equation} 
\begin{equation}\begin{aligned}
&\mathbb{E}\left[\left|\boldsymbol{g}_{E}^{T}\text{diag}\left(\boldsymbol{p}_{k}\right)\boldsymbol{\tilde{h}}_{k}^{(2)}\right|^{2}\right]\\
&=\mathbb{E}\left[\boldsymbol{\bar{g}}_{E}^{T}\boldsymbol{\Gamma}^{\frac{1}{2}}\text{diag}\left(\boldsymbol{p}_{k}\right)\mathbb{E}\left[\boldsymbol{\tilde{h}}_{k}^{(2)}\boldsymbol{\tilde{h}}_{k}^{(2)^{H}}\right]\text{diag}\left(\boldsymbol{p}_{k}\right)\boldsymbol{\Gamma}^{\frac{1}{2}}\boldsymbol{\bar{g}}_{E}^{*}\right]\\
&=\text{tr}\left(\boldsymbol{\Gamma}^{\frac{1}{2}}\text{diag}\left(\boldsymbol{p}_{k}\right)\boldsymbol{R}_{k}^{(2)}\text{diag}\left(\boldsymbol{p}_{k}\right)\boldsymbol{\Gamma}^{\frac{1}{2}}\right)=\text{tr}\left(\boldsymbol{p}_{k}^{T}\boldsymbol{\Gamma}\boldsymbol{R}_{k}^{(2)}\boldsymbol{p}_{k}\right)\\
&=\text{tr}\left(\boldsymbol{P}_{k}\boldsymbol{\Gamma}\boldsymbol{R}_{k}^{(2)}\right)=\text{tr}\left(\boldsymbol{P}_{k}\boldsymbol{\bar{B}}_{k}\right)\label{ApAc4b}
\end{aligned}\end{equation}  
\noindent
where $\boldsymbol{\tilde{h}}_{k}^{(2)}=\boldsymbol{\hat{h}}_{k}^{*}-\tau\sqrt{\phi P_E}\boldsymbol{C}_{k}\boldsymbol{g}_{E}^{*}$ and $\boldsymbol{R}_{k}^{(2)}=\mathbb{E}[\boldsymbol{\tilde{h}}_{k}^{(2)}\boldsymbol{\tilde{h}}_{k}^{(2)^{H}}]=\boldsymbol{\bar{R}}_{k}-\tau^{2}\phi P_{E}\boldsymbol{C}_{k}^{2}\boldsymbol{\Gamma}$. The third equality in \eqref{ApAc3} is obtained by substituting the value of $\boldsymbol{w}_{k}$ from \eqref{eeq4a}, \eqref{eeq2a} and \eqref{eeq2c}. The fourth equality in \eqref{ApAc3} follows from the statistical independence between $\boldsymbol{g}_{E}$ and $\boldsymbol{\tilde{h}}_{k}^{(2)}$. The first equality in \eqref{ApAc4a} follows from applying Corollary 1 in \cite{evans2000large} and the diagonality of the matrices $\boldsymbol{\Gamma}$, $\boldsymbol{C}_{k}$ and $\text{diag}\left(\boldsymbol{p}_{k}\right)$. In the first equality in \eqref{ApAc4b}, the expectation is moved to $\boldsymbol{\tilde{h}}_{k}^{(2)}\boldsymbol{\tilde{h}}_{k}^{(2)^{H}}$ based on the statistical independence between $\boldsymbol{\bar{g}}_{E}$ and $\boldsymbol{\tilde{h}}_{k}^{(2)}$. The second equality in \eqref{ApAc4b} follows from applying Corollary 1 in \cite{evans2000large}. The third and the fourth equalities in in \eqref{ApAc4b} follow from the diagonality of the matrices $\boldsymbol{\Gamma}$, $\boldsymbol{R}_{k}^{(2)}$ and $\text{diag}\left(\boldsymbol{p}_{k}\right)$. By substituting \eqref{ApAc4a} and \eqref{ApAc4b} in \eqref{ApAc3}, we get $\mathbb{E}[X_k]=\mathbb{E}[|b_{k}|^2]=\tau^{2}\phi P_{E}d_{k}^{2}+d_{k}^{(1)}$, $d_{k}^{2}=\text{tr}(\boldsymbol{P}_{k}\boldsymbol{B}_{k})$ and $d_{k}^{(1)}=\text{tr}(\boldsymbol{P}_{k}\boldsymbol{\bar{B}}_{k})$. 

For $\mathbb{E}[Y_{k}]$, based on the statistical independence between $\boldsymbol{\bar{w}}_{k}$ and $\boldsymbol{w}$, we have
\begin{equation}\begin{aligned}\label{ApAc05}
\mathbb{E}\left[Y_{k}\right]&=\mathbb{E}\left[|\boldsymbol{g}_{E}^{T}(\boldsymbol{\bar{w}}_{k}z+\boldsymbol{w})|^{2}+\bar{n}\right]\\
&=\mathbb{E}\left[|\boldsymbol{g}_{E}^{T}\boldsymbol{\bar{w}}_{k}|^{2}\right]+\mathbb{E}\left[|\boldsymbol{g}_{E}^{T}\boldsymbol{w}|^{2}\right]+\sigma_{n}^{2}.
\end{aligned}\end{equation} 
\noindent
\begin{equation}\begin{aligned}\label{Apxxc05}
&\mathbb{E}\left[|\boldsymbol{g}_{E}^{T}\boldsymbol{\bar{w}}_{k}|^{2}\right]\\
&=\tau^{2}\phi P_{E}\mathbb{E}\left[|\boldsymbol{g}_{E}^{T}\text{diag}\left(\boldsymbol{\bar{p}}\right)\boldsymbol{C}_{k}\boldsymbol{g}_{E}^{*}|^{2}\right]+\mathbb{E}\left[|\boldsymbol{g}_{E}^{T}\text{diag}\left(\boldsymbol{\bar{p}}\right)\boldsymbol{\tilde{h}}_{k}^{(2)^{*}}|^{2}\right]\\
&=\tau^{2}\phi P_{E}|\text{tr}\left(\text{diag}\left(\boldsymbol{\bar{p}}\right)\boldsymbol{\Gamma}\boldsymbol{C}_{k}\right)|^{2}\\
&+\mathbb{E}\left[\boldsymbol{\bar{g}}_{E}^{T}\boldsymbol{\Gamma}^{\frac{1}{2}}\text{diag}\left(\boldsymbol{\bar{p}}\right)\mathbb{E}\left[\boldsymbol{\tilde{h}}_{k}^{(2)^{*}}\boldsymbol{\tilde{h}}_{k}^{(2)^{T}}\right]\text{diag}\left(\boldsymbol{\bar{p}}\right)\boldsymbol{\Gamma}^{\frac{1}{2}}\boldsymbol{\bar{g}}_{E}^{*}\right]\\
&=\tau^{2}\phi P_{E}|\boldsymbol{\bar{p}}^{T}\text{diag}\left(\boldsymbol{\Gamma}\boldsymbol{C}_{k}\right)|^{2}+\boldsymbol{\bar{p}}^{T}\boldsymbol{\Gamma}\boldsymbol{R}_{k}^{(2)}\boldsymbol{\bar{p}}\\
&=\tau^{2}\phi P_{E}\text{tr}\left(\boldsymbol{\bar{P}}\text{diag}\left(\boldsymbol{\Gamma}\boldsymbol{C}_{k}\right)\text{diag}\left(\boldsymbol{\Gamma}\boldsymbol{C}_{k}\right)
^{H}\right)+\text{tr}\left(\boldsymbol{\bar{P}}\boldsymbol{\Gamma}\boldsymbol{R}_{k}^{(2)}\right)\\
&=\tau^{2}\phi P_{E}\text{tr}(\boldsymbol{\bar{P}}\boldsymbol{B}_{k})+\text{tr}(\boldsymbol{\bar{P}}\boldsymbol{\bar{B}}_{k})=\tau^{2}\phi P_{E}\bar{d}_{k}^{2}+\bar{d}_{k}^{(1)}.
\end{aligned}\end{equation} 
The first equality in \eqref{Apxxc05} follows from substituting the value of $\boldsymbol{\bar{w}}_{k}$ from \eqref{eeq4b}, \eqref{eeq2a} and \eqref{eeq2c}; and the statistical independence between $\boldsymbol{g}_{E}$ and $\boldsymbol{\tilde{h}}_{k}^{(2)}$. The first term in the second equality follows from applying Corollary 1 in \cite{evans2000large}. In the second term after the second equality, the expectation is moved to $\boldsymbol{\tilde{h}}_{k}^{(2)}\boldsymbol{\tilde{h}}_{k}^{(2)^{H}}$ based on the statistical independence between $\boldsymbol{\bar{g}}_{E}$ and $\boldsymbol{\tilde{h}}_{k}^{(2)}$. The first term in the third equality follows from the diagonality of the matrices $\boldsymbol{\Gamma}$, $\boldsymbol{C}_{k}$ and $\text{diag}\left(\boldsymbol{\bar{p}}_{k}\right)$, while the second term follows since the entries of $\boldsymbol{\bar{g}}_{E}$ are zero mean unit variance independent RVs and the matrices $\boldsymbol{\Gamma}$ and $\boldsymbol{R}_{k}^{(2)}$ are diagonal. The fourth equality follows since the matrices $\boldsymbol{\Gamma}$, $\boldsymbol{C}_{k}$ and $\boldsymbol{R}_{k}^{(2)}$ are diagonal.
\noindent
\begin{equation}\begin{aligned}\label{Apjxc05}
&\mathbb{E}\left[|\boldsymbol{g}_{E}^{T}\boldsymbol{w}|^{2}\right]=\mathbb{E}\left[\boldsymbol{\bar{g}}_{E}^{T}\boldsymbol{\Gamma}^{\frac{1}{2}}\text{diag}\left(\boldsymbol{p}\right)\mathbb{E}\left[\boldsymbol{\hat{g}}^{*}\boldsymbol{\hat{g}}^{T}\right]\text{diag}\left(\boldsymbol{p}\right)\boldsymbol{\Gamma}^{\frac{1}{2}}\boldsymbol{\bar{g}}_{E}^{*}\right]\\
&=\boldsymbol{p}^{T}\boldsymbol{\Gamma}\boldsymbol{R}~\boldsymbol{p}=\text{tr}(\boldsymbol{P}\boldsymbol{B})=d.
\end{aligned}\end{equation}
\noindent
The first equality in \eqref{Apjxc05}, the expectation is moved to $\boldsymbol{\hat{g}}^{*}\boldsymbol{\hat{g}}^{T}$ based on the statistical independence between $\boldsymbol{\bar{g}}_{E}$ and $\boldsymbol{\hat{g}}$. The second equality follows since the entries of $\boldsymbol{\bar{g}}_{E}$ are zero mean unit variance independent RVs. The third equality follows since the matrices $\boldsymbol{\Gamma}$ and $\boldsymbol{R}$ are diagonal. By substituting the results from \eqref{Apxxc05} and \eqref{Apjxc05} in \eqref{ApAc05} we get
\begin{equation}\begin{aligned}\label{ApAcxs05}
 &\mathbb{E}\left[Y_k\right]=\mathbb{E}\left[|\hat{b}_{k}|^{2}+|b|^{2}+\sigma_{n}^{2}\right]=\tau^{2}\phi P_{E}\bar{d}_{k}^{2}+\bar{d}_{k}^{(1)}+d+\sigma_{n}^{2}, 
\end{aligned}\end{equation}  
\noindent
 where $d=\text{tr}(\boldsymbol{P}\boldsymbol{B}),~\bar{d}_{k}^{2}=\text{tr}(\boldsymbol{\bar{P}}\boldsymbol{B}_{k})$ and $\bar{d}_{k}^{(1)}=\text{tr}(\boldsymbol{\bar{P}}\boldsymbol{\bar{B}}_{k})$. For $\text{var}(Y_{k})$ we have 
{\fontsize{9}{10}\begin{equation}\begin{aligned}\label{ApAddc5}
&\text{var}(Y_{k})=\mathbb{E}\left[\left|Y_{k}-\mathbb{E}[Y_k]\right|^{2}\right]=\mathbb{E}\left[\left|\left|\boldsymbol{g}_{E}^{T}\boldsymbol{w}\right|^{2}-\text{tr}\left(\boldsymbol{P}\boldsymbol{B}\right)\right|^{2}\right]\\
&+\mathbb{E}\left[\left|\left|\boldsymbol{g}_{E}^{T}\boldsymbol{\bar{w}}_{k}\right|^{2}-\tau^{2}\phi P_{E}\text{tr}\left(\boldsymbol{\bar{P}}\boldsymbol{B}_{k}\right)-\text{tr}\left(\boldsymbol{\bar{P}}\boldsymbol{\bar{B}}_{k}\right)\right|^{2}\right]+\sigma_{n}^{2},
\end{aligned}\end{equation} }
\noindent
Now let us calculate the first and the second terms in \eqref{ApAddc5}, as follows
{\fontsize{9}{10}\begin{equation}\begin{aligned}\label{ApAc6a}
&\mathbb{E}\left[\left|\left|\boldsymbol{g}_{E}^{T}\boldsymbol{w}\right|^{2}-\text{tr}\left(\boldsymbol{P}\boldsymbol{B}\right)\right|^{2}\right]\\
&=\mathbb{E}\left[\left|\boldsymbol{g}_{E}^{T}\boldsymbol{w}\right|^{4}-2\left|\boldsymbol{g}_{E}^{T}\boldsymbol{w}\right|^{2}\text{tr}\left(\boldsymbol{P}\boldsymbol{B}\right)+\text{tr}^{2}\left(\boldsymbol{P}\boldsymbol{B}\right)\right]\\
&=\mathbb{E}\left[\left|\boldsymbol{g}_{E}^{T}\boldsymbol{w}\right|^{4}\right]-\text{tr}^{2}\left(\boldsymbol{P}\boldsymbol{B}\right)~~~~~~~~~~~~~~\\
&=\mathbb{E}\left[\left(\sum_{\mathbb{I}}\bar{g}_{E_j}\sqrt{\gamma_{j}p_{j}}\hat{g}_{j}^{*}\hat{g}_{m}\sqrt{\gamma_{m}p_{m}}\bar{g}_{E_m}^{*}\right)^{2}\right]~~~~~~~~~~~~~~~~\\
&+\mathbb{E}\left[\left(\sum_{j}\left|\bar{g}_{E_j}\right|^{2}\gamma_{j}p_{j}\left|\hat{g}_{j}\right|^{2}\right)^{2}\right]-\text{tr}^{2}\left(\boldsymbol{P}\boldsymbol{B}\right)\\
&=2~\text{tr}^{2}\left(\boldsymbol{P}\boldsymbol{B}\right)-\text{tr}\left(\left(\boldsymbol{P}\boldsymbol{B}\right)^{\circ 2}\right).
\end{aligned}\end{equation} }
{\fontsize{9}{10}\begin{equation}\begin{aligned}
&\mathbb{E}\left[\left|\left|\boldsymbol{g}_{E}^{T}\boldsymbol{\bar{w}}_{k}\right|^{2}-\tau^{2}\phi P_{E}\text{tr}\left(\boldsymbol{\bar{P}}\boldsymbol{B}_{k}\right)-\text{tr}\left(\boldsymbol{\bar{P}}\boldsymbol{\bar{B}}_{k}\right)\right|^{2}\right]\\
&=\mathbb{E}\left[\left|\left|\boldsymbol{g}_{E}^{T}\text{diag}\left(\boldsymbol{\bar{p}}\right)\boldsymbol{\tilde{h}}_{k}^{(2)}\right|^{2}-\text{tr}\left(\boldsymbol{\bar{P}}\boldsymbol{\bar{B}}_{k}\right)\right|^{2}\right]\\
&=\mathbb{E}\left[\left|\boldsymbol{g}_{E}^{T}\text{diag}\left(\boldsymbol{\bar{p}}\right)\boldsymbol{\tilde{h}}_{k}^{(2)}\right|^{4}\right]-\text{tr}^{2}\left(\boldsymbol{\bar{P}}\boldsymbol{\bar{B}}_{k}\right)\\
&=\mathbb{E}\left[\left(\sum_{\mathbb{I}}\bar{g}_{E_j}\sqrt{\gamma_{j}\bar{p}_{j}}\left[\hat{h}_{k}^{(2)}\hat{h}_{k}^{(2)^{H}}\right]_{j,m}\sqrt{\gamma_{m}\bar{p}_{m}}\bar{g}_{E_m}^{*}\right)^{2}\right]~~~~~~~\\
&+\mathbb{E}\left[\left(\sum_{j}\left|\bar{g}_{E_j}\right|^{2}\gamma_{j}p_{j}\left[\hat{h}_{k}^{(2)}\hat{h}_{k}^{(2)^{H}}\right]_{j,j}\right)^{2}\right]-\text{tr}^{2}\left(\boldsymbol{\bar{P}}\boldsymbol{\bar{B}}_{k}\right)\\
&=2~\text{tr}^{2}\left(\boldsymbol{\bar{P}}\boldsymbol{\bar{B}}_{k}\right)-\text{tr}\left(\left(\boldsymbol{\bar{P}}\boldsymbol{\bar{B}}_{k}\right)^{\circ 2}\right),\label{ApAvvc6b}
\end{aligned}\end{equation}}
\noindent  
where $\circ$ denotes the Hadamard power operation and $\mathbb{I}=\{\{j\}\times\{m\}|~j\neq m\}$. The second equality in \eqref{ApAddc5} follows from the statistical independence between $\boldsymbol{w}$, $\boldsymbol{\bar{w}}_{k}$ and $\bar{n}$. The second equality in \eqref{ApAc6a} is obtaioned by substituting the value of $\mathbb{E}[|\boldsymbol{g}_{E}^{T}\boldsymbol{w}|^{2}]$ from \eqref{Apjxc05}. The second equality in \eqref{ApAvvc6b} is obtained since the norm $\left|\boldsymbol{g}_{E}^{T}\boldsymbol{\bar{w}}_{k}\right|^{2}$ converges to $\tau^{2}\phi P_{E}\text{tr}(\boldsymbol{\bar{P}}\boldsymbol{B}_{k})+|\boldsymbol{g}_{E}^{T}\text{diag}(\boldsymbol{\bar{p}})\boldsymbol{\tilde{h}}_{k}^{(2)}|^{2}$ in the signal domain.  By expanding the norms $|\boldsymbol{g}_{E}^{T}\boldsymbol{w}|^{4}$ and $|\boldsymbol{g}_{E}^{T}\text{diag}(\boldsymbol{\bar{p}})\boldsymbol{\tilde{h}}_{k}^{(2)}|^{4}$ -- which are composed of squared exponential RVs -- followed by applying the statistical expectation\footnote{The expectation is obtained by applying the fact: For an exponential RV $X\sim \mathcal{E}(\lambda)$, the $n$th moment of $X$ is $\mathbb{E}[X^{n}]=\frac{\Gamma(n+1)}{\lambda^{n}}=\frac{\Gamma(n+1)}{\lambda^{n}}=\frac{n!}{\lambda^{n}}$.}, we obtain the final results in \eqref{ApAc6a} and \eqref{ApAvvc6b}, respectively. 

Given that the entries of the matrix $\boldsymbol{B}_{k}$ have non-zero positive values, the matrices $\boldsymbol{\bar{B}}_{k}$ and $\boldsymbol{B}$ are diagonal, and the additive terms in $(\text{tr}(\boldsymbol{\bar{P}}\boldsymbol{B}_{k}) + \text{tr}(\boldsymbol{\bar{P}}\boldsymbol{\bar{B}}_{k}) + \text{tr}(\boldsymbol{P}\boldsymbol{B}))^{2}$ are of a finite order of magnitude; then, asymptotically, we have $(\mathbb{E}[Y_{k}])^{2}\overset{N\to \infty}{\to}\mathcal{O}\left(N^{4}\right)$ and $\text{var}(Y_{k})\overset{N\to \infty}{\to}\mathcal{O}\left(2N^{2}\right)$. This implies that $\frac{\text{var}(Y_k)}{(E[Y_k])^2}\overset{N\to \infty}{\to}0$ and $R = f(\text{var}(Y_k), \text{cov}(X_k, Y_k))\overset{N\to \infty}{\to}0$. Based on \eqref{ApAc2} and this result, and since $\text{cov}(X_k, Y_k) = 0$ (follows from the statistical independence between $X_k$ and $Y_k$), we have  
\begin{equation}\begin{aligned}\label{ApAc7}
  &\mathbb{E}\left[\text{SINR}_{E_k}\right]=\\
  &\mathbb{E}\left[\frac{X_k}{Y_k}\right]\overset{N\to \infty}{\to}\frac{\mathbb{E}[X_k]}{\mathbb{E}[Y_k]}=\frac{\tau^{2}\phi P_{E}d_{k}^{2}+d_{k}^{(1)}}{d+\tau^{2}\phi P_{E}\bar{d}_{k}^{2}+\bar{d}_{k}^{(1)}+\sigma_{n}^{2}}.
\end{aligned}\end{equation}
\noindent
By substituting \eqref{ApAc7} in \eqref{ApAc1}, we get \eqref{eeq13rr2}--\eqref{eeq20}. This concludes the proof. 

\subsection{Deriving the asymptotic value of AHE in \eqref{eeq22}} \label{AHE_Ana} 
The details of deriving $\mathbb{E} [|b_{k}|^{2}]$ and $\mathbb{E} [|\hat{b}_{ k}|^{2}+|b|^{2}]$ are provide \eqref{ApAc3}-\eqref{ApAc4b} and \eqref{ApAc05}-\eqref{ApAcxs05}, respectively. The details of deriving the values $\mathbb{E} [|b_{j\neq k}|^{2}]$, $\mathbb{E} [|\tilde{b}_{j}|^{2}]$, $\mathbb{E} [|\tilde{\hat{b}}_{j}|^{2}]$ and $\mathbb{E} [|\tilde{b}|^{2}]$ that constitute $\bar{E}_{k}$ in \eqref{eeq22} are as follows

{\fontsize{10}{10}\begin{equation}\begin{aligned}\label{Apjfgc05}
&\mathbb{E} \left[|b_{j\neq k}|^{2}\right]=\mathbb{E}\left[|\boldsymbol{g}_{E}^{T}\boldsymbol{w}_{j\neq k}|^{2}\right]=\mathbb{E}\left[\boldsymbol{\bar{g}}_{E}^{T}\boldsymbol{\Gamma}^{\frac{1}{2}}\text{diag}\left(\boldsymbol{p}_{j}\right)\mathbb{E}\left[\boldsymbol{\hat{h}}_{j}^{*}\boldsymbol{\hat{h}}_{j}^{T}\right]\right.\\
&\left.\text{diag}\left(\boldsymbol{p}_{j}\right)\boldsymbol{\Gamma}^{\frac{1}{2}}\boldsymbol{\bar{g}}_{E}^{*}\right]=\boldsymbol{p}_{j}^{T}\boldsymbol{\Gamma}\boldsymbol{R}_{j}~\boldsymbol{p}_{j}=\text{tr}(\boldsymbol{P}_{j}\boldsymbol{\tilde{B}}).
\end{aligned}\end{equation}}
{\fontsize{10}{10}\begin{equation}\begin{aligned}\label{Apjxchj5}
&\mathbb{E} \left[|\tilde{b}_{j}|^{2}\right]=\mathbb{E}\left[|\boldsymbol{g}^{T}\boldsymbol{w}_{j}|^{2}\right]=\mathbb{E}\left[\boldsymbol{\bar{g}}^{T}\boldsymbol{\Gamma}^{\frac{1}{2}}\text{diag}\left(\boldsymbol{p}_{j}\right)\mathbb{E}\left[\boldsymbol{\hat{h}}_{j}^{*}\boldsymbol{\hat{h}}_{j}^{T}\right]\right.\\
&\left.\text{diag}\left(\boldsymbol{p}_{j}\right)\boldsymbol{\Gamma}^{\frac{1}{2}}\boldsymbol{\bar{g}}^{*}\right]=\boldsymbol{p}_{j}^{T}\boldsymbol{\Gamma}\boldsymbol{R}_{j}~\boldsymbol{p}_{j}=\text{tr}(\boldsymbol{P}_{j}\boldsymbol{\tilde{B}}),
\end{aligned}\end{equation}}
{\fontsize{10}{10}\begin{equation}\begin{aligned}\label{Ap89chj5}
&\mathbb{E} \left[|\tilde{\hat{b}}_{k}|^{2}\right]=\mathbb{E}\left[|\boldsymbol{g}^{T}\boldsymbol{\bar{w}}_{k}|^{2}\right]=\mathbb{E}\left[\boldsymbol{\bar{g}}^{T}\boldsymbol{\Gamma}^{\frac{1}{2}}\text{diag}\left(\boldsymbol{\bar{p}}\right)\mathbb{E}\left[\boldsymbol{\hat{h}}_{k}^{*}\boldsymbol{\hat{h}}_{k}^{T}\right]\right.\\
&\left.\text{diag}\left(\boldsymbol{\bar{p}}\right)\boldsymbol{\Gamma}^{\frac{1}{2}}\boldsymbol{\bar{g}}^{*}\right]=\boldsymbol{\bar{p}}^{T}\boldsymbol{\Gamma}\boldsymbol{\bar{R}}_{k}~\boldsymbol{\bar{p}}=\text{tr}(\boldsymbol{P}_{j}\boldsymbol{\tilde{B}}_{k}).
\end{aligned}\end{equation}}
\noindent
In the second equalities in \eqref{Apjfgc05}, \eqref{Apjxchj5} and \eqref{Ap89chj5}, the expectation is moved to $\boldsymbol{\hat{h}}_{j\neq k}^{*}\boldsymbol{\hat{h}}_{j\neq k}^{T}$, $\boldsymbol{\hat{h}}_{j}^{*}\boldsymbol{\hat{h}}_{j}^{T}$ and $\boldsymbol{\hat{h}}_{k}^{*}\boldsymbol{\hat{h}}_{k}^{T}$ based on the statistical independence between $\boldsymbol{\bar{g}}_{E}$ and $\boldsymbol{\hat{h}}_{j\neq k}$, $\boldsymbol{\bar{g}}$ and $\boldsymbol{\hat{h}}_{j}$, and between $\boldsymbol{\bar{g}}_{E}$ and $\boldsymbol{\hat{h}}_{k}$, respectively. The third equalities in \eqref{Apjfgc05}, \eqref{Apjxchj5} and \eqref{Ap89chj5} follows since the entries of $\boldsymbol{\bar{g}}_{E}$ and $\boldsymbol{\bar{g}}$ are zero mean unit variance independent RVs. The fourth equalities follows since the matrices $\boldsymbol{\Gamma}$, $\boldsymbol{R}_{j}$ and $\boldsymbol{\bar{R}}_{k}$ are diagonal. \\

\begin{equation}\begin{aligned}\label{Aprss1}
&\mathbb{E} \left[|\tilde{b}|^{2}\right]=\mathbb{E}\left[|\boldsymbol{g}^{T}\boldsymbol{w}|^{2}\right]=\mathbb{E}\left[|\boldsymbol{g}^{T}\text{diag}\left(\boldsymbol{p}\right)\boldsymbol{C}\left(\tau\sqrt{(1-\phi)P_{E}}\boldsymbol{g}^{*}\right.\right.\\
&\left.\left.+\boldsymbol{N}^{*}\psi_{E}\right)|^{2}\right]=\tau^{2}(1-\phi)P_{E}\mathbb{E}\left[|\boldsymbol{g}^{T}\text{diag}\left(\boldsymbol{p}\right)\boldsymbol{C}\boldsymbol{g}^{*}|^{2}\right]+\\
&\mathbb{E}\left[|\boldsymbol{g}^{T}\text{diag}\left(\boldsymbol{p}\right)\boldsymbol{C}\boldsymbol{N}^{*}\psi_{E}|^{2}\right]\\
&=\tau^{2}(1-\phi)P_{E}|\text{tr}\left(\text{diag}\left(\boldsymbol{p}\right)\boldsymbol{C}\boldsymbol{\Gamma}\right)|^{2}\\
&+\mathbb{E}\left[\boldsymbol{\bar{g}}^{T}\boldsymbol{\Gamma}^{\frac{1}{2}}\text{diag}\left(\boldsymbol{p}\right)\boldsymbol{C}\mathbb{E}\left[\boldsymbol{N}^{*}\psi_{E}\psi_{E}^{H}\boldsymbol{N}^{T}\right]\boldsymbol{C}\text{diag}\left(\boldsymbol{p}\right)\boldsymbol{\Gamma}^{\frac{1}{2}}\boldsymbol{\bar{g}}^{*}\right]\\
&=\tau^{2}(1-\phi)P_{E}|\boldsymbol{p}^{T}\text{diag}\left(\boldsymbol{C}\boldsymbol{\Gamma}\right)|^{2}+\tau\sigma_{n}^{2}\boldsymbol{p}^{T}\boldsymbol{\Gamma}\boldsymbol{C}^{2}\boldsymbol{p}\\
&=\tau^{2}(1-\phi)P_{E}\tilde{d}^{2}+\tau\sigma_{n}^{2} \tilde{d}^{(1)}\\
&=\tau^{2}(1-\phi)P_{E}\text{tr}\left(\boldsymbol{P}\boldsymbol{\ddot{B}}\right)+\tau\sigma_{n}^{2}\text{tr}\left(\boldsymbol{P}\boldsymbol{\hat{B}}\right).
\end{aligned}\end{equation}
\noindent
The second equality follows from the statistical independent between $\boldsymbol{g}$ and $\boldsymbol{N}$. The first term after the fourth equality follows from applying Corollary 1 in \cite{evans2000large}. In the second term in the fourth equality, the expectation is moved to $\boldsymbol{N}^{*}\psi_{E}\psi_{E}^{H}\boldsymbol{N}^{T}$ based on the statistical independence between $\boldsymbol{\bar{g}}$ and $\boldsymbol{N}$. The first term in the fifth equality follows since the matrices $\boldsymbol{\Gamma}$, $\boldsymbol{C}$ and $\text{diag}\left(\boldsymbol{p}\right)$ are diagonal. The second term in the fifth equality follows since $\mathbb{E}[\boldsymbol{N}^{*}\psi_{E}\psi_{E}^{H}\boldsymbol{N}^{T}]=\tau\sigma_{n}^{2} \boldsymbol{I}_{N}$ and the entries of $\boldsymbol{\bar{g}}$ are zero mean unit variance independent RVs. The seventh equality  follows since the matrices $\boldsymbol{\Gamma}$ and $\boldsymbol{C}$ are diagonal. The form which is after the fifth equality is identical to that used in \eqref{eeq22}, while the SDP form which is after the seventh equality is identical to that used in \eqref{CFeeq30e}.  
\section{Proof of Theorem \ref{theorem_3}} \label{Appen_A0} 
To prove that the optimal solution $\{\boldsymbol{P}_{i}^{\star}\},~\boldsymbol{\bar{P}}^{\star},~\boldsymbol{P^{\star}}$ obtained by solving \eqref{CFeeq30} is always of unity rank, we exploit the boundedness property of the dual Lagrangian function to show that the optimal primal matrices $\{\boldsymbol{P}_{i}^{\star}\},~\boldsymbol{\bar{P}}^{\star},~\boldsymbol{P^{\star}}$ can satisfy the KKT conditions of optimality at one case in which $\{\text{rank}(\boldsymbol{P}_{i}^{\star})\},~\text{rank}(\boldsymbol{\bar{P}}^{\star}),~\text{rank}(\boldsymbol{P}^{\star})=1$, and that has been validated by computer simulation. 

The Lagrangian of the equivalent problem \eqref{CFeeq31} is 
\begin{flalign}\label{ApA1}
&\mathcal{L}\left(\mathbb{S}, \mathbb{L} \right)=\sum_{k}\text{tr}\left(\boldsymbol{P}_{k}\boldsymbol{\Pi}_{k}\right)+\text{Tr}\left(\boldsymbol{\bar{P}}\boldsymbol{\bar{\Pi}}\right)+\text{Tr}\left(\boldsymbol{P}\boldsymbol{\Pi}\right)+\bar{d},
\end{flalign}

\noindent
where

\noindent
\begin{flalign}\label{ApA2}
\boldsymbol{\Pi}_{k}=&\lambda_{2_k}\tau^{2}P_{I}\boldsymbol{A}_{k}+\sum_{j\neq k}\lambda_{2_j}\boldsymbol{A}_{j,k}-\sum_{j\neq k}\lambda_{3_j}\boldsymbol{A}_{j,k}+(\lambda_{6_k}\zeta\nonumber\\
&-\lambda_{4_k})\left(\tau^{2}\phi P_{E}\boldsymbol{B}_{k}+\boldsymbol{\bar{B}}_{k}\right)+\lambda_{6_j}\zeta\bigg(\sum_{j\neq k}\boldsymbol{\tilde{B}}_{k}+\sum_{j}\boldsymbol{\tilde{B}}_{k}\bigg)\nonumber\\
&-\sum_{l}\bigg(\lambda_{7_k}\boldsymbol{D}_{l}\boldsymbol{\bar{R}}_{k}+\sum_{j\neq k}\lambda_{7_j}\boldsymbol{D}_{l}\boldsymbol{R}_{k}\bigg)+\boldsymbol{F}_{k},
\end{flalign}
\begin{flalign}\label{ApA3}
\boldsymbol{\bar{\Pi}}=&\sum_{k}\bigg[(\lambda_{2_k}-\lambda_{3_k})\left(\tau^{2}P_{I}\boldsymbol{A}_{k}+\boldsymbol{\bar{A}}_{k}\right)+(\lambda_{5_k}-\lambda_{4_k})\nonumber\\
&\left(\tau^{2}P_{I}\boldsymbol{A}_{k}+\boldsymbol{\bar{A}}_{k}\right)+\lambda_{6_k}\zeta\left(\tau^{2}\phi P_{E}\boldsymbol{B}_{k}+\boldsymbol{\bar{B}}_{k}+\boldsymbol{\tilde{B}}_{k}\right)\nonumber\\
&+\sum_{l}\lambda_{7_k}\boldsymbol{D}_{l}\boldsymbol{\bar{R}}_{k}\bigg]+\boldsymbol{\bar{F}},
\end{flalign}
\begin{flalign}\label{ApA4}
\boldsymbol{\Pi}=&\sum_{k}\bigg[(\lambda_{2_k}-\lambda_{3_k})\boldsymbol{\tilde{A}}_{k}+(\lambda_{5_k}-\lambda_{4_k})\boldsymbol{B}+\lambda_{6_k}\zeta\left(\boldsymbol{B}+\boldsymbol{\hat{B}}\right.\nonumber\\
&\left.+\tau^{2}(1-\phi) P_{E}\boldsymbol{\ddot{B}}\right)\bigg]+\boldsymbol{F},
\end{flalign}
\begin{flalign}\label{ApA5}
\bar{d}=&\pi+\sum_{k=1}\bigg[\lambda_{1_k}\left(u_{k}-s_{k}-t_{k}+v_{k}-\pi\right)-\lambda_{2_k}e^{u_{k}}+\lambda_{3_k}e^{s_{k}}\nonumber\\
&+\lambda_{4_k}e^{t_{k}}-\lambda_{5_k}e^{v_{k}}+\left(\lambda_{2_k}-\lambda_{3_k}\right)\sigma_{n}^{2}\frac{\tau_{d}+1}{\tau_d}\nonumber\\
&+\left(\lambda_{5_k}-\lambda_{4_k}\right)\sigma_{n}^{2}-\lambda_{6_k}\bar{E}+\lambda_{7_k}P_{T}\bigg].
\end{flalign}

\noindent
$\mathbb{L}=\{\{\lambda_{1_k}\},~\dots,~ \{\lambda_{7_k}\},~\{\boldsymbol{F}_{k}\},~\boldsymbol{\bar{F}},~\boldsymbol{F}\}$ are the Lagrange multipliers of the constraints (\ref{CFeeq31a}), (\ref{CFeeq30a})--(\ref{CFeeq30f}) and the constraints on $\{\boldsymbol{P}_{k}\}$, $\boldsymbol{\bar{P}}$ and $\boldsymbol{P}$ in  (\ref{CFeeq30g}), respectively, with $\{\lambda_{j_k}\}\geq 0$, $\{\boldsymbol{F}_{k}\}, ~\boldsymbol{\bar{F}},~ \boldsymbol{F}\succeq \boldsymbol{0}$. Now, for the Lagrangian function to exist, the infimum of $\mathcal{L}$ over the primal variable $\mathbb{S}$, $\inf_{\mathbb{S}}\mathcal{L}$, should be bounded from below, therefore, we have  
\begin{flalign}\label{ApA6}
&\boldsymbol{\Pi}_{k},~\boldsymbol{\bar{\Pi}},~\boldsymbol{\Pi}\succeq \boldsymbol{0},~\left\{\lambda_{3_k},~\lambda_{4_k}\right\}=0,~\left\{\lambda_{1_k}\right\}\geq 1,~\forall k.
\end{flalign}

Given that Slater's condition holds (see \eqref{CFeeq32} and the paragraph that follows) and based on the non-negativeness of the dual variables (Lagrange multipliers), the satisfaction of the KKT's complementary slackness condition results in $\text{tr}(\boldsymbol{P}_{k}^{\star}\boldsymbol{F}_{k}^{\star})=0~\forall~k$. The KKT's stationarity condition should satisfy $\sum_{k}\frac{\partial \mathcal{L}}{\partial \boldsymbol{P}_{k}^{\star}}=\boldsymbol{0}$, therefore
\begin{flalign}\label{ApA7}
\sum_{k}\boldsymbol{\Pi}_{k}^{\star}=\boldsymbol{0}.
\end{flalign} 

\noindent
Given the fact that if the summation of multiple positive semi-definite matrices is equal to zero, all the matrices are equal to zero. And based on the \eqref{ApA2}, \eqref{ApA6} and \eqref{ApA7}, we have    
\begin{flalign}\label{ApA8}
\boldsymbol{F}_{k}^{\star}=&-\lambda_{2_k}\tau^{2}P_{I}\boldsymbol{A}_{k}+\boldsymbol{G}_{k}^{\star},
\end{flalign}
\begin{flalign}\label{ApA9}
&\boldsymbol{G}_{k}^{\star}=-\sum_{j\neq k}\lambda_{2_j}^{\star}\boldsymbol{A}_{j,k}-\lambda_{6_k}^{\star}\zeta\left(\tau^{2}\phi P_{E}\boldsymbol{B}_{k}+\boldsymbol{\bar{B}}_{k}\right)-\lambda_{6_j}^{\star}\zeta\nonumber\\
&\bigg(\sum_{j\neq k}\boldsymbol{\tilde{B}}_{k}+\sum_{j}\boldsymbol{\tilde{B}}_{k}\bigg)+\sum_{l}\bigg(\lambda_{7_k}^{\star}\boldsymbol{D}_{l}\boldsymbol{\bar{R}}_{k}+\sum_{j\neq k}\lambda_{7_j}^{\star}\boldsymbol{D}_{l}\boldsymbol{R}_{k}\bigg).
\end{flalign}

Let $\text{null}\left(\boldsymbol{F}_{k}^{\star}\right)=\boldsymbol{\Omega}_{k}=\left[\boldsymbol{\omega}_{k,1}, ..., \boldsymbol{\omega}_{k,N-\text{rank}(\boldsymbol{F}_{k}^{\star})}\right]$ and $\text{null}\left(\boldsymbol{G}_{k}^{\star}\right)=\boldsymbol{\Psi}_{k}=\left[\boldsymbol{\psi}_{k,1}, ..., \boldsymbol{\psi}_{k,N-\text{rank}(\boldsymbol{G}_{k}^{\star})}\right]$. By making use of the inequality of matrix sum \cite[subsection 3.3.4]{MatrixAlgebra2017} and \eqref{ApA8} we have\footnote{Please note that $\text{rank}(\boldsymbol{A}_{k})=1$, this is understandable from the structure of $\boldsymbol{A}_{k}$. Please refer to the first paragraph in \ref{Prob_Formu}.} $\text{rank}(\boldsymbol{F}_{k}^{\star})\geq\text{rank}(\boldsymbol{G}_{k}^{\star})-1 $. Based on this result, and since $\text{rank}(\boldsymbol{\Omega}_{k})=N-\text{rank}(\boldsymbol{F}_{k}^{\star})$ and $\text{rank}(\boldsymbol{\Psi}_{k})=N-\text{rank}(\boldsymbol{G}_{k}^{\star})$, then the following result is true 
\begin{equation}\begin{aligned}\label{ApA11}
 \text{rank}\left(\boldsymbol{\Omega}_{k}\right)\leq \text{rank}\left(\boldsymbol{\Psi}_{i}\right)+1.
\end{aligned}\end{equation}

Now, let us examine the null space of $\boldsymbol{G}_{k}^{\star}$, $\boldsymbol{\psi}_{k,j}\in\boldsymbol{\Psi}_{k}$, by computing the inner product between $\boldsymbol{\psi}_{k,j}$ and $\boldsymbol{F}_{k}^{\star}$ in \eqref{ApA8} as follows
\begin{flalign}\label{ApA12}
 \boldsymbol{\psi}_{k,j}^{H}\boldsymbol{F}_{k}^{\star}\boldsymbol{\psi}_{k,j}=-\boldsymbol{\psi}_{k,j}^{H}\left(\lambda_{2_k}^{\star}\tau^{2}P_{I}\boldsymbol{A}_{k}\right)\boldsymbol{\psi}_{k,j}\leq 0,
\end{flalign}

\noindent
where the inequality in \eqref{ApA12} follows from\footnote{$\boldsymbol{A}_{k}\succeq 0$ follows since it is structured from a vector whose all entries are positive (see \eqref{CFeeq29} and the paragraph that follows).} $\boldsymbol{A}_{k}\succeq 0$. However, since $\boldsymbol{F}_{k}^{\star}\succeq 0$, \eqref{ApA12} can only hold with equality, i.e., $\boldsymbol{\psi}_{k,j}^{H}\left(\lambda_{2_k}^{\star}\tau^{2}P_{I}\boldsymbol{A}_{k}\right)\boldsymbol{\psi}_{k,j}= 0$. This result implies that the null space of $\boldsymbol{G}_{k}^{\star}$ always forms null space of $\boldsymbol{F}_{k}^{\star}$, i.e., $\boldsymbol{\Psi}_{k}$ is a sub-matrix of $\boldsymbol{\Omega}_{k}$, therefore, and according to \eqref{ApA11}, $\boldsymbol{\omega}_{k,j}\in\boldsymbol{\Omega}_{k}$ belongs to one of the following two spaces: 1) the column space of $\boldsymbol{\Psi}_{k}$, $\boldsymbol{\omega}_{k,j}\in\{\boldsymbol{\psi}_{k,j}\}$; 2) 1-dimensional vector space, $\boldsymbol{\omega}_{k,j}=\boldsymbol{a}\in\mathcal{C}^{N\times 1}$ where $\boldsymbol{a}\notin\{\boldsymbol{\psi}_{k,j}\}$.    

Since the optimal value of $\boldsymbol{P}_{k}^{\star}$ needs to satisfy the complementary slackness condition, $\text{tr}(\boldsymbol{P}_{k}^{\star}\boldsymbol{F}_{k}^{\star})=0~\forall~k$, the structure of $\boldsymbol{P}_{k}^{\star}$ is 
\begin{flalign}\label{ApA13}
\boldsymbol{P}_{k}^{\star}=\sum_{i=1}^{L\leq N}m_{k,j}\boldsymbol{q}_{j}\boldsymbol{q}_{j}^{H},~~\boldsymbol{q}_{j}\in\{\boldsymbol{\psi}_{k,j},~\boldsymbol{a}\},
\end{flalign}

\noindent
where $\{m_{k,j}\}$ are non-negative scaling factors. The $\boldsymbol{P}_{k}^{\star}$'s component $m_{k,j}\boldsymbol{\psi}_{k,j}\boldsymbol{\psi}_{k,j}^{H}$ introduces zero information signal power at $\text{IU}_k$ since $\boldsymbol{\psi}_{k,j}^{H}\boldsymbol{A}_{k}\boldsymbol{\psi}_{k,j}= 0$, and therefore contributes by a negative ESR. Thus, $m_{k,j}\boldsymbol{\psi}_{k,j}\boldsymbol{\psi}_{k,j}^{H}$ is a non-optimal component of $\boldsymbol{P}_{k}^{\star}$. By this, we can conclude that $\boldsymbol{P}_{k}^{\star}$ is constructed by the single component $\boldsymbol{P}_{k}^{\star}=m_{k,1}\boldsymbol{a}\boldsymbol{a}^{H}$, $\boldsymbol{a}\notin\{\boldsymbol{\psi}_{k,j}\}$, therefore, $\text{rank}(\boldsymbol{P}_{k}^{\star})=1$ is always true. This concludes the proof.

%
\section{Proofs of Lemmas \ref{lemma_3} and \ref{lemma_4}} \label{Appen_C} 

\subsection{Proof of Lemma \ref{lemma_3}}

Since the exponential function is convex (has a downward curvature), the tangent line at any point is below the function trajectory. Using triangulation (as depicted in Fig. \ref{Apfig2}), it can be easily understood that the value of the Taylor first order approximation of $e^{x}$, $e^{\bar{x}}(x-\bar{x}+1)$, always lies at the tangent line ($L_{1}$, black solid line) which is always below the function trajectory (circle-marked line). Fig. \ref{Apfig2} shows the case $\bar{x}<x$. Following a comparable reasoning, the previous result can be proved for the other case $\bar{x}>x$. This concludes the proof. 

\subsection{Proof of Lemma \ref{lemma_4}}

\begin{figure}[t]
\begin{center}
\captionsetup{justification=centering}
\includegraphics[width=0.4\textwidth,trim = 1cm 0cm 1cm 2cm, clip]{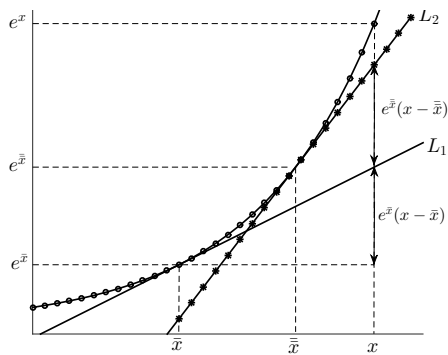}
\caption{The geometry of the successive first order approximation.}
\label{Apfig2}
\end{center}
\end{figure}

 Building upon the proof of Lemma \ref{lemma_3}, the value of successive Taylor approximation $e^{\bar{\bar{x}}}(x-\bar{\bar{x}}+1)$, $e^{\bar{\bar{x}}}=e^{\bar{x}}(x-\bar{x}+1)$, lies at the tangent line ($L_{2}$, star-marked line in Fig. \ref{Apfig2}) touched at $(\bar{\bar{x}},e^{\bar{\bar{x}}})$. Since the derivative of the exponential function is non-decreasing, $L_{2}$ always lies above $L_{1}$ for $x\geq \bar{\bar{x}}$. Therefore, $e^{\bar{\bar{x}}}(x-\bar{\bar{x}}+1)>e^{\bar{x}}(x-\bar{x}+1)$ is always true. This concludes the proof.

\ifCLASSOPTIONcaptionsoff
  \newpage
\fi

\end{document}